\documentclass[useAMS,usenatbib]{mn2e}
\usepackage{epsfig,natbib}
\usepackage{eufrak}
\usepackage{amsmath}
\usepackage{natbib}
\usepackage{graphicx}
\usepackage{color}
\usepackage{rotating}

\title[FOF Groups and Clusters in 2SLAQ]
{Friends-of-Friends Groups and Clusters in the 2SLAQ Catalogue}
\author[S.~Farrens et al.]
{\parbox[t]{\textwidth}{S. Farrens$^{1,2}$\thanks{Email: farrens@ieec.uab.es}, F.B. Abdalla$^2$,  E.S. Cypriano$^3$, C. Sabiu$^2$ \& C. Blake$^4$}\vspace*{6pt}\\
$^1$Institut de Ci\`{e}ncies de lÕEspai, IEEC-CSIC, Campus UAB, F. de Ci\`{e}ncies, Torre C5 par-2, Barcelona 08193, Spain\\
$^2$Department of Physics \& Astronomy, University College London, Gower Street, London WC1E 6BT, UK.\\
$^3$Departamento de Astronomia, IAG-USP, Rua do Mat\~{a}o 1226, 05508-090, S\~{a}o Paulo, Brazil.\\
$^4$Centre for Astrophysics \& Supercomputing, Swinburne University of Technology, P.O. Box 218, Hawthorn, VIC 3122, Australia.}

\date{Accepted ???, Received ???; in original form \today}
\pagerange{\pageref{firstpage}--\pageref{lastpage}}
\pubyear{2008}
\begin{document}

\maketitle
\label{firstpage}

\begin{abstract}
We present  a catalogue of galaxy groups and clusters selected using a friends-of-friends algorithm with a dynamic linking length from the 2dF-SDSS and QSO (2SLAQ) luminous red galaxy survey. The linking parameters for the code are chosen through an analysis of simulated 2SLAQ haloes. The resulting catalogue includes 313 clusters containing 1,152 galaxies. The galaxy groups and clusters have an average velocity dispersion of $\overline{\sigma}_{v}=467.97$ kms$^{-1}$ and an average size of $\overline{R}_{clt}=0.78$ Mpc $h^{-1}$. Galaxies from regions of one square degree and centred on the galaxy clusters were downloaded from the Sloan Digital Sky Survey Data Release 6 (SDSS DR6). Investigating the photometric redshifts and cluster red-sequence of these galaxies shows that the galaxy clusters detected with the FoF algorithm are reliable out to z$\sim$0.6. We estimate masses for the clusters using 
their velocity dispersions. These mass estimates are shown to be consistent with 2SLAQ mock halo masses. Further analysis of the simulation haloes shows that clipping out low richness groups with large radii improves the purity of catalogue from 52$\%$ to 88$\%$, while retaining a completeness of 94$\%$. Finally, we test the two-point correlation function of our cluster catalogue. We find a best-fitting power law model , $\xi(r)=(r/r_0)^{\gamma}$, with parameters $r_0=24 \pm 4$ Mpc $h^{-1}$ and $\gamma=-2.1 \pm 0.2$, which are in agreement with other low redshift cluster samples and consistent with a $\Lambda$CDM universe.
\end{abstract}


\section{Introduction}
Galaxy clusters are the largest bound objects in the Universe and are important structures for examining the distribution of matter  and how this evolves with time. Investigating how the mass-function of galaxy clusters changes with time provides an effective method for constraining cosmological parameters \citep{PressSchechter:74,Peebles:93,Weller:02,Weller:03,Haiman:01}. 

The first galaxy cluster was unknowingly detected by Charles Messier, who recorded the positions of 11 nebulae in the Virgo cluster in the 18$^{\textrm{th}}$ century. Later evidence for the existence of clusters of galaxies was provided in the work of Harlow Shapley and Adelaide Ames in the 1930Õs \citep{ShapleyAmes:32}. They severely undermined the idea that galaxies are randomly distributed throughout the Universe. Probably the most significant figure in the pioneering of galaxy cluster detection was George Abell \citep{Abell:58}. Abell surveyed around three quarters of the sky using photographic plates, which meant that he had to identify the locations of galaxy overdensities by eye. To avoid including field galaxies, Abell defined galaxy clusters as regions 1.5 Mpc $h^{-1}$ in radius (the Abell Radius) containing fifty or more galaxies that are no more than two magnitudes fainter than the third brightest member of the group. Remarkably, later studies have confirmed a large number of Abell's clusters as being genuine bound structures. Due to the simplicity of this two-dimensional approach, however, serious problems can arise from \emph{e.g.} inhomogeneities, photometric errors and projection effects. \citet{Bahcall:83} found an excess of power in the angular correlation function using Abell clusters and \citet{Vanhaarlem:97} show, using N-body simulations, that $\sim1/3$ of Abell's clusters have incorrect richness estimates and $\sim1/3$ of Abell richness class $R\geq1$ clusters are missed. Around thirty years after Abell's work hybrid photo-digital surveys were able to improve upon the photographic plate method for detecting clusters of galaxies \citep{Maddox:90}. It was, however, the advent of digital CCD surveys that brought about significant advances, a good example being surveys such as the SDSS \citep{York:00}, 2dFGRS \citep{Colless:99}, 6dF \citep{Jones:06} and the ongoing GAMA \citep{Driver:08}. Large sky surveys like SDSS signify a major step forward in obtaining galaxy data, however analysing that data can be approached and interpreted in many different ways. Automated algorithms supply a means of reducing subjectivity in the analysis of large data sets. Finally, the availability of increasingly detailed simulations in recent years has enabled more powerful tests of the completeness and reliability of cluster catalogues.

In the last few decades many different algorithms have been developed to find galaxy clusters. The \emph{Counts in Cells} method \citep{Couch:91,LidmanPeterson:96} looks for enhancements of galaxy surface density in a given area. \emph{Percolation} methods  group together galaxies that are separated by a distance on the sky less than a given threshold distance \citep{Efstathiou:88,Davis:85,Dalton:97,Ramella:02}. This technique was originally applied to redshift surveys by \citet{HuchraGeller:82}. An extended version of Huchra and Geller's friend-of-friends algorithm was later developed to deal with photometric redshifts by \citet{Botzler:04}. The changes implemented in the extended friend-of-friends algorithm were necessary to account for the large errors associated with photometric redshifts \citep{Botzler:04}. \emph{Matched Filter} techniques model the spatial and luminosity distributions of galaxies in a cluster to generate a cluster likelihood map \citep{Postman:96,Kim:02}. \emph{Voronoi Tessellation} decomposes the galaxies in a region of space into discrete points surrounded by cells facilitating the identification of overdensities \citep{Kim:02,Lopes:04,Ramella:01}. Another method is to look for a \emph{Red Sequence} in the colour-magnitude relation of galaxy clusters \citep{GladdersYee:00,Koester:07}. The \emph{Cut and Enhance} method makes colour and colour-colour cuts to produce subsamples of galaxies in different redshift ranges \citep{Goto:02}. A review of various methods and techniques used for the optical detection of galaxy clusters is provided by \citet{Gal:06}.

The Percolation or friend-of-friends (FoF) algorithm is the cluster finding method of interest for the purposes of this paper. The 2dF-SDSS and QSO (2SLAQ) Luminous Red Galaxy Survey, which is subset of photometrically selected luminous red galaxies (LRGs) from SDSS \citep{Cannon:06}, is a good starting place for the use of the FoF algorithm because of its relatively small size and the abundance of both spectroscopic and photometric data available.

This paper is the first in a series of two papers exploring the detection and analysis of galaxy clusters in the 2SLAQ catalogue and will focus on utilising the spectroscopic data available in the catalogue. The second paper, Farrens et al. (in prep), will investigate detecting clusters using photometric redshifts estimated from the SDSS photometry in the 2SLAQ catalogue and compare the results with those provided in this paper. This comparison will demonstrate the reliability of the FoF technique to detect clusters with photometric redshifts on a consistent data set. Large data sets are required if one is to do cosmology with galaxy clusters. The fastest way of obtaining this data is through photometric surveys. Upcoming surveys like the Dark Energy Survey (DES), Euclid and Planck will obtain such photometric data for large numbers of galaxies across the sky and since it is not feasible to obtain spectra for all of these objects within the observing time frames, it is therefore necessary to develop reliable methods for detecting clusters using photometric data. Photometric cluster catalogues made from surveys such as these can cover a larger volume than spectroscopic catalogues and can thus be used to probe the shape of the mass function at higher masses.

The following section provides some background on the 2SLAQ survey. Section \ref{sec:fofs} describes the Huchra and Geller Friends-of-friends method in detail and how we implemented the algorithm. Section \ref{sec:mock} provides a description of the mock catalogue used to calibrate the linking parameters. Section \ref{sec:datas} presents the resulting groups and clusters found using the FoF algorithm and the analysis made on these results. Finally, section \ref{sec:2pcfs} shows the 2-point correlation function of the clusters. We present a sample of the cluster catalogue in the appendix of this paper.

\section{The 2dF-SDSS and QSO (2SLAQ) Luminous Red Galaxy Survey}
\label{sec:2slaqs}
2SLAQ is a spectroscopic survey of around 15000 potential luminous red galaxies (LRGs) in the redshift range 0.45 $\leq$ z $\leq$ 0.7 \citep{Cannon:06}. The target LRGs were selected photometrically from the Sloan Digital Sky Survey \citep{York:00}. 2SLAQ also includes a lower resolution survey of faint quasi-stellar objects (QSOs) or quasars. Observations began in March 2003 using the Two-degree Field instrument (2dF) on the 3.9m Anglo-Australian Telescope. 

The 2SLAQ criteria for selecting galaxies from the Sloan Digital Sky Survey consisted in placing limits on the magnitudes, colours and star-galaxy separations of the SDSS data. The magnitude limits were imposed by requiring that:
\begin{equation}\label{eq:e}
17.5 \ {\leq} \ i_{deV}-A_{i}<19.8 \; 
\end{equation}
\begin{equation}\label{eq:f}
i_{fibre}<21.2 \; 
\end{equation}

\noindent where $i_{deV}$ is the total magnitude based on a fit of each galaxy to a de Vaucouleurs profile, $A_i$ is the extinction in the $i$-band and $i_{fibre}$ is the flux contained within the aperture of a spectroscopic fibre in the \emph{i}-band. These limits enabled the selection of bright LRGs out to $z\sim0.8$ and eliminated objects too diffuse to produce useful spectra. The colour cuts were such that:
\begin{equation}\label{eq:g}
0.5<g-r<3.0 \; 
\end{equation}
\begin{equation}\label{eq:h}
r-i<2.0 \; 
\end{equation}
\begin{equation}\label{eq:i}
c_{\parallel}{\equiv}0.7(g-r)+1.2(r-i-0.18)>1.6 \; 
\end{equation}
\begin{equation}\label{eq:j}
d_{\perp}{\equiv}(r-i)-\frac{(g-r)}{8.0}>0.5 \; 
\end{equation}

\noindent where $c_{\parallel}$ eliminates later-type galaxies and $d_{\perp}$ selects early-type galaxies at increasingly high redshift. Equations \ref{eq:g} and \ref{eq:h} ensure that only objects near the main locus of LRGs are selected. The star-galaxy separation criteria was:
\begin{equation}\label{eq:k}
i_{psf}-i_{model}>0.2+0.2(20.0-i_{deV}) \; 
\end{equation}
\begin{equation}\label{eq:l}
\textrm{radius}_{deV(i)}>0.2"§ \; 
\end{equation}

\noindent where $i_{psf}$ is the \emph{i}-band magnitude of an isolated star fitted with a point spread function (PSF) model, $i_{model}$ is the model \emph{i}-band magnitude, which is a good proxy for the PSF, and radius$_{deV}$ is the de Vaucouleurs radius, which is the effective radius in the model magnitude. These limits eliminate the majority of stellar contamination.

\section{Friends-of-Friends Method}
\label{sec:fofs}
\subsection{The Huchra and Geller Friends-of-Friends Algorithm}
The Huchra and Geller friends-of-friends algorithm was developed to search for groups of galaxies in the magnitude limited CFA1 redshift survey \citep{HuchraGeller:82,Huchra:83,Geller:83}.  This technique is commutative and utilises only a galaxy's right ascension, declination and redshift to detect structure by finding galaxies that are separated by a distance less than some threshold, D$_L$, and that have a velocity difference less than some threshold, V$_L$. The threshold values are chosen depending on the properties of the galaxies in the catalogue. For simplicity $D_L$ and $V_L$ can be set as fixed values, however this may lead to some selection effects being ignored. \citet{HuchraGeller:82} adopt a method that compensates for the variation in the completeness of the galaxy luminosity function as a function of redshift.

\subsection{Choice of Dynamic Friends-of-Friends Linking}
\begin{figure}
	\centering
	\includegraphics[width=6cm]{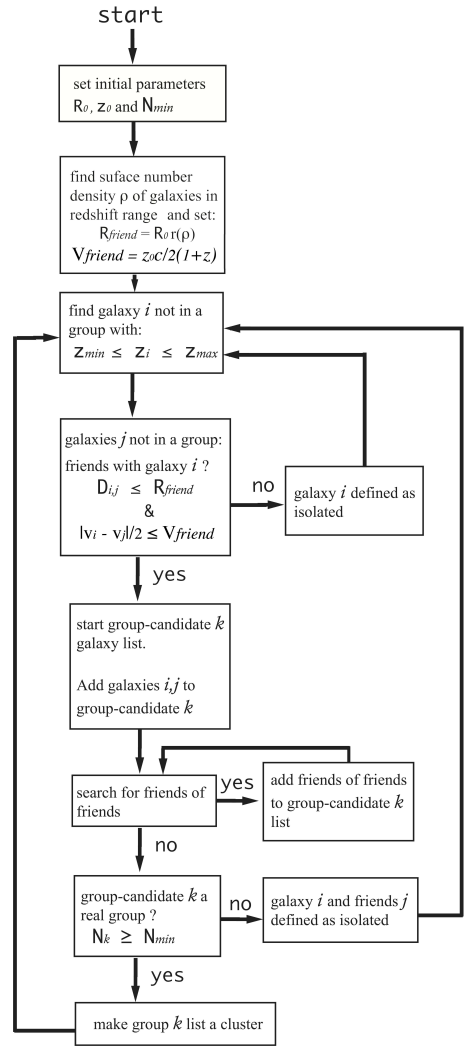}  
	\caption{{Fluxogram of the FoF algorithm with dynamic linking length. Based on \citet{HuchraGeller:82}.}\label{fig:fof0}}
\end{figure}

The friends-of-friends algorithm used in this paper follows the \citet{HuchraGeller:82} method in most respects. The only difference lies in the linking parameters, which change with the surface number density of galaxies at a given redshift. This dynamic linking length is more appropriate to this sample of galaxies as is does not assume a magnitude limit pre-selection, which would remove large numbers of galaxies usable for cluster detection and prevent any clusters at the highest redshifts from being found. The choice of this type of linking parameter has the drawback of allowing for a more complex selection function that will render interpretation less straight forward. A fluxogram of the Dynamic Friends-of-Friends (hereafter DFoF) algorithm is shown in fig.\ref{fig:fof0}. 

Following the steps shown in fig.\ref{fig:fof0}: First the initial values of linking length, R$_0$, redshift linking, $z_0$, and the minimum number of galaxy members needed to form a group/cluster, N$_{min}$, are chosen. To compensate for selection effects, R$_{0}$ is varied via equation \ref{eq:qop}:
\begin{equation}\label{eq:qop}
R_{friend}(z)\propto R_0\bigg(\frac{dN}{dz}\frac{dz}{dV}\frac{1}{A_{sky}}\bigg)^{-\frac{1}{2}} \; 
\end{equation}

\noindent where $z$ is redshift of each galaxy for which the variable linking length is calculated, $dN(z)\//dz$ is the surface number density of galaxies in the redshift range covered by the galaxy catalogue ($z_{\textrm{min}}\leq z\leq z_{\textrm{max}}$), $dV\//dz$ is the differential comoving volume and $A_{sky}$ is the fraction of sky area covered in the catalogue relative to the total sky area. Note that the limits $z_{\textrm{min}}$ and $z_{\textrm{max}}$ can be modified if one does not wish to use the full redshift range of the catalogue. $z_0$ is converted into a  velocity linking parameter, $v_{friend}(z)$, via equation \ref{eq:voz}:
\begin{equation}\label{eq:voz}
v_{friend}(z)=\frac{z_0c}{2(1+z)} \; 
\end{equation}

\noindent where $c$ is the speed of light. The factor of 2 in the denominator accounts for the fact that we want to be able to link galaxies that have velocities approaching along the line of sight to those that have velocities receding along the line of sight. $v_{friend}(z)$ will therefore be representative of the velocity dispersions of the clusters being examined.

A galaxy $i$ is selected from the catalogue that has not yet been assigned to a group and lies in the redshift range $z_{min}\leq z\leq z_{max}$. The projected distance between galaxy $i$ and second galaxy $j$, $D_{ij}$, is calculated using:
\begin{equation}\label{eq:q}
D_{ij}=cos^{-1}(sin(\delta_i)sin(\delta_j)+cos(\delta_i)cos(\delta_j)cos(\alpha_i-\alpha_j)) \; 
\end{equation}
\noindent where $\alpha$ and $\delta$ are right ascension and declination respectively.

The two galaxies are linked together (\emph{i.e.} are friends) if they satisfy the conditions in equations \ref{eq:r} and \ref{eq:s}:
\begin{equation}\label{eq:r}
D_{ij}\leq R_{friend}(z) \; 
\end{equation}
\begin{equation}\label{eq:s}
\frac{|v_i-v_j|}{2}\leq v_{friend}(z) \; 
\end{equation}

A group-candidate $k$ is formed that includes galaxy $i$ and its friends. A search is then made around the galaxies linked to $i$. This process is repeated until no further friends are found. The group-candidate $k$ is defined as a real group if it satisfies equation \ref{eq:t}:
\begin{equation}\label{eq:t}
N_k\geq N_{min} \; 
\end{equation}

\noindent For the purposes of this paper $N_{min} \geq 3$. This is a reasonable assumption when dealing with LRGs, which are not common objects.

\section{Mock Catalogue}
\label{sec:mock}
To determine the optimum values of R$_0$ and z$_0$ for the DFoF code, a mock galaxy catalogue that simulates the 2SLAQ catalogue was produced. This mock catalogue contains a distribution of 7,651,076 dark matter haloes and 824,704 galaxies across an octant of the sky.

The halo catalogue was derived from the Horizon $4\pi$ simulation \citep{teyssier:09,prunet:08}. This is a $\Lambda$CDM dark matter $\it{N}$-body simulation using WMAP 3 cosmology with a $2$h$^{-1}$ Gpc periodic box on a grid of 40,963 cells. The $7 \times 10^{10}$ particles were evolved using the Particle Mesh scheme of the RAMSES code \citep{teyssier:02} on an adaptively refined grid (AMR) with around $1.4 \times 10^{11}$ cells, reaching a formal resolution of 262,144 cells in each direction ($\sim$ 7h$^{-1}$ kpc comoving). The simulation covers a sufficiently large volume to compute a full-sky dark matter distribution, while resolving Milky-Way size haloes with more than 100 particles and exploring small scales deeply into the non-linear regime. The dark matter distribution in the simulation was integrated in a light cone out to redshift 1, around an observer located at the centre of the simulation box. The underlying cosmology for WMAP 3 is: $\Omega_M$ = 0.24, $\Omega_\Lambda$ = 0.76, $\Omega_b$ = 0.042, n = 0.958, H$_0$ = 73 and $\sigma_8$ = 0.77.

Gravitationally bound haloes of dark matter are selected using the spherical overdensity HOP\footnote{http://cmb.as.arizona.edu/$\sim$eisenste/hop/hop.html} method of \citet{Eisenstein:98}. HOP is based on a hybrid approach in which the local density field is first obtained by smoothing the density field with an SPH-like kernel using the $n$ nearest neighbours. Then the particles above a given threshold are linked with their highest density neighbours until, after several ``hops'', they are connected to the one particle with the highest density within the region above the threshold. All particles linked to the local density maximum are identified as a group. 

The haloes were then populated using the Halo Occupation Distribution (HOD), where the number of galaxies residing	within each halo is drawn from a probability, $P(N |M )$, that a dark matter halo of mass $M$ will host $N$ galaxies.

The first moment of $P(N|M)$ is the mean number of galaxies as a function of halo mass and it is usually parameterised as a sum of a central and a satellite components \citep{Kravtsov:04, Tinker:07}. 

The probability that a halo contains a central galaxy is given by: 
\begin{equation}
\langle N_c | M \rangle = 0.5 \left[1+ {\rm erf}\left(\frac{\log_{10}(M/M_{cut})}{\sigma_{cut}}\right)\right]
\end{equation}
and the number of satellite galaxies is obtained from a Poisson sampling of:
\begin{equation}
\langle N_s | M \rangle = \left(\frac{M}{M_0}\right)^{\beta}
\end{equation}

\begin{figure}
	\centering
	\includegraphics[width=8cm]{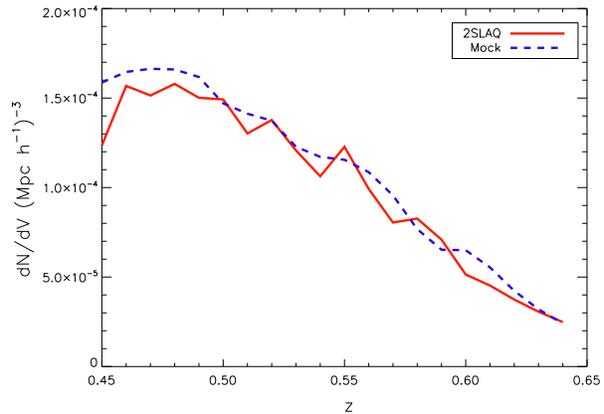}   
	\caption{{Number density as a function of redshift for the mock halo catalogue (blue dashed line) and the 2SLAQ catalogue (red solid line).}\label{fig:massf}}
\end{figure}

The HOD model used is that of \citet{blake:08} and includes several derived parameters that were computed at various redshifts bins between $0.4<z<0.7$. Table \ref{tab:hod} lists the HOD parameters in each redshift bin. For the purposes of this paper an evolving HOD model is constructed that smoothly interpolates between the four \citet{blake:08} redshift bins as shown in fig.\ref{fig:massf}. This figure shows that our smoothly evolving HOD model, based on \citet{blake:08}, succeeds in reproducing the observed number density of the 2SLAQ data. 

\begin{table}
\caption The HOD parameters derived from the MegaZ-LRG sample using the methodology of \citet{blake:08}. The small differences between these parameters and those of \citet{blake:08} reflect the difference in the assumed underlying cosmology.
\begin{center}
\begin{tabular}{c c c c c}
\hline
Redshift slice & $\sigma_{cut}$ & $\log\left(\frac{M_0}{M_\odot/h}\right)$ & $\beta$ & $\log\left(\frac{M_{cut}}{M_\odot/h}\right)$ \\ 
\hline
\hline
$0.45 < z < 0.50$	& 0.618	& 13.88	& 1.41 & 12.96\\
$0.50 < z < 0.55$	& 0.469	& 13.99	& 1.54 & 13.00\\
$0.55 < z < 0.60$	& 0.554	& 14.16	& 1.66 & 13.23\\
$0.60 < z < 0.65$	& 0.675	& 14.43	& 1.56 & 13.60\\
\hline
\end{tabular}
\end{center}
\label{tab:hod}
\end{table}
The radial positions of satellite galaxies within a halo are assigned according to the NFW profile \citep{NFW:96}. Specifically,  it is assumed that the mass inside a given radius, properly normalised, represents the probability of containing a galaxy. Thus, integrating to the halo boundary would yield a probability of one and therefore all galaxies would be placed within this radius. The angular position relative to the halo centre is chosen randomly for each galaxy. 

In our model, all dark matter particles are assumed to be in approximately spherical virialised haloes.  The velocity
of a dark matter particle is the sum of two terms,
\begin{equation}
v = v_{\rm vir} + v_{\rm halo},
\label{eq:vsum}
\end{equation}
the first is due to the velocity of the particle about the centre of mass of its parent halo, and the second is due to the
motion of the centre of mass of the parent.

Consider the first term, $v_{\rm vir}$. We will assume that virialised haloes are isothermal spheres, so that the distribution
of velocities within them is Maxwellian. This is in reasonable agreement with measurements of virial velocities within haloes
in numerical simulations. If $\sigma_{vir}$ denotes the rms speeds of particles within a halo, then the virial theorem requires
that:
\begin{equation}
{Gm\over r}\propto \sigma_{vir}^2 \propto
{H(z)^2\over 2}\,\Delta^{1/3}_{vir}(z)
\left(3m\over 4\pi\rho_{crit}(z)\right)^{2/3} \,
\label{eq:vcirc}
\end{equation}
where the final proportionality comes from the fact that all haloes have the same density whatever their mass:
$m/r^3\propto \Delta_{vir}\,\rho_{crit}$.  This shows that $\sigma_{vir}\propto m^{1/3}$:  the more massive haloes are expected to be `hotter'.  At fixed mass, the constant of proportionality depends on time and cosmology, and on the exact shape of the density profile of the halo.  A convenient fitting formula is provided by \cite{Bryan:98}:
\begin{equation}
\sigma_{\rm vir}(m,z) = 102.5\,g_\sigma\, \Delta^{1/6}_{\rm vir}(z)\,
\left(H(z)\over H_0\right)^{1/3}
\left({m\over 10^{13} M_\odot/h}\right)^{1/3}\,
\label{eq:sigmavir}
\end{equation}
where $g_\sigma = 0.9$, and
\begin{equation}
\Delta_{\rm vir} = 18\pi^2 + 82 x - 39 x^2, \qquad {\rm with}\quad
             x = \Omega(z)-1  \,
\end{equation}
and $\Omega(z) = [\Omega_m\,(1+z)^3]\, [H_0/H(z)]^2$.
It has been shown that $\sigma(M)$ is independent of local environment \citep{Sheth:01}, however it may depend on position within a halo. This will mainly be due to the fact that haloes have complicated density and velocity profiles.

\begin{figure}
	\centering
	\includegraphics[width=8cm]{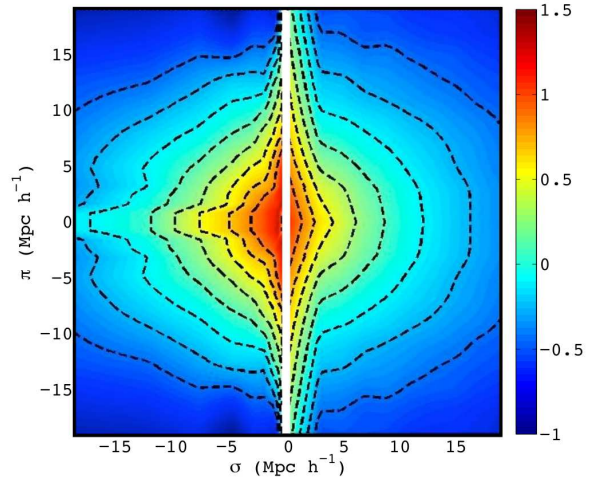}  
	\caption{{The two-dimensional redshift-space correlation function for the 2SLAQ catalogue (left half) and the mock catalogue (right half) plotted as a function of transverse, $\sigma$, and radial, $\pi$, pair separation. The colour map represents $\log(\xi)$. The 2SLAQ correlation function is that of \citet{Ross:07}.}\label{fig:corrf}}
\end{figure}

The two-dimensional redshift-space correlation function for the mock catalogue, $\xi(\sigma,\pi)$, was calculated with the pair separation decomposed in terms of perpendicular, $\sigma$, and parallel, $\pi,$ distances. Each component of separation was calculated for 20 bins equally spaced in comoving distance between $0.01 < \sigma, \pi [{\rm Mpc} h^{-1}]< 40$, creating a 400 element grid of measurements. The smooth contours are created by linear interpolation over the grid. Fig.\ref{fig:corrf} shows the calculated 2-D correlation functions for the 2SLAQ catalogue of \citet{Ross:07} and the mock galaxy catalogue. It shows the expected linear squashing at large radii and redshift distortions with `finger of God' elongations at small scales.  The colour map represents $\log(\xi)$. 

Fig.\ref{fig:massf} and \ref{fig:corrf} show that the simulation reproduces the properties of 2SLAQ well enough for us to believe that it is sufficient to aid us in calibrating our cluster finding parameters and therefore any limits imposed upon the values of  R$_0$ and z$_0$ from the simulation will be applicable to the real data. The discrepancy between the real and mock 2SLAQ data in fig.\ref{fig:corrf} at low values of $\pi$ is due to shot noise (\emph{i.e.} there is a small number of pairs at this scale). There are, however, some caveats that should be taken into account that may cause the simulation to not be fully representative of reality. For example, the effects of things such as the way in which galaxies are added to dark matter haloes and cluster mergers at high redshifts could have influence on the results. In the following section, we discuss how the simulations can bias the results together with the analysis of the results.

\section{Data Analysis}
\label{sec:datas}
\subsection{Linking Parameter Optimisation}
The DFoF code was run on the 2SLAQ simulation with different values of R$_0$ and z$_0$ to find the optimum combination of the two. To gain a more physical interpretation of these parameters, hereafter R$_0$ and z$_0$ are expressed in terms of $R_{friend}(z)$ and $v_{friend}(z)$ at $z=0.5$ using equations \ref{eq:qop} and \ref{eq:voz}. The parameters were varied in the ranges:
\begin{equation}
 0 \leq R_{friend}(z=0.5) \leq 4.2\;(\textrm{Mpc}\;h^{-1}) \;
\end{equation} 
\begin{equation}
100 \leq v_{friend}(z=0.5) \leq 1500 \:\textrm{(kms}^{-1}\textrm{)}\;
\end{equation} 
The choice of parameters is made based on the completeness, purity and total number of clusters in the resulting catalogues.
\begin{figure*}
	\centering
	\includegraphics[width=8cm]{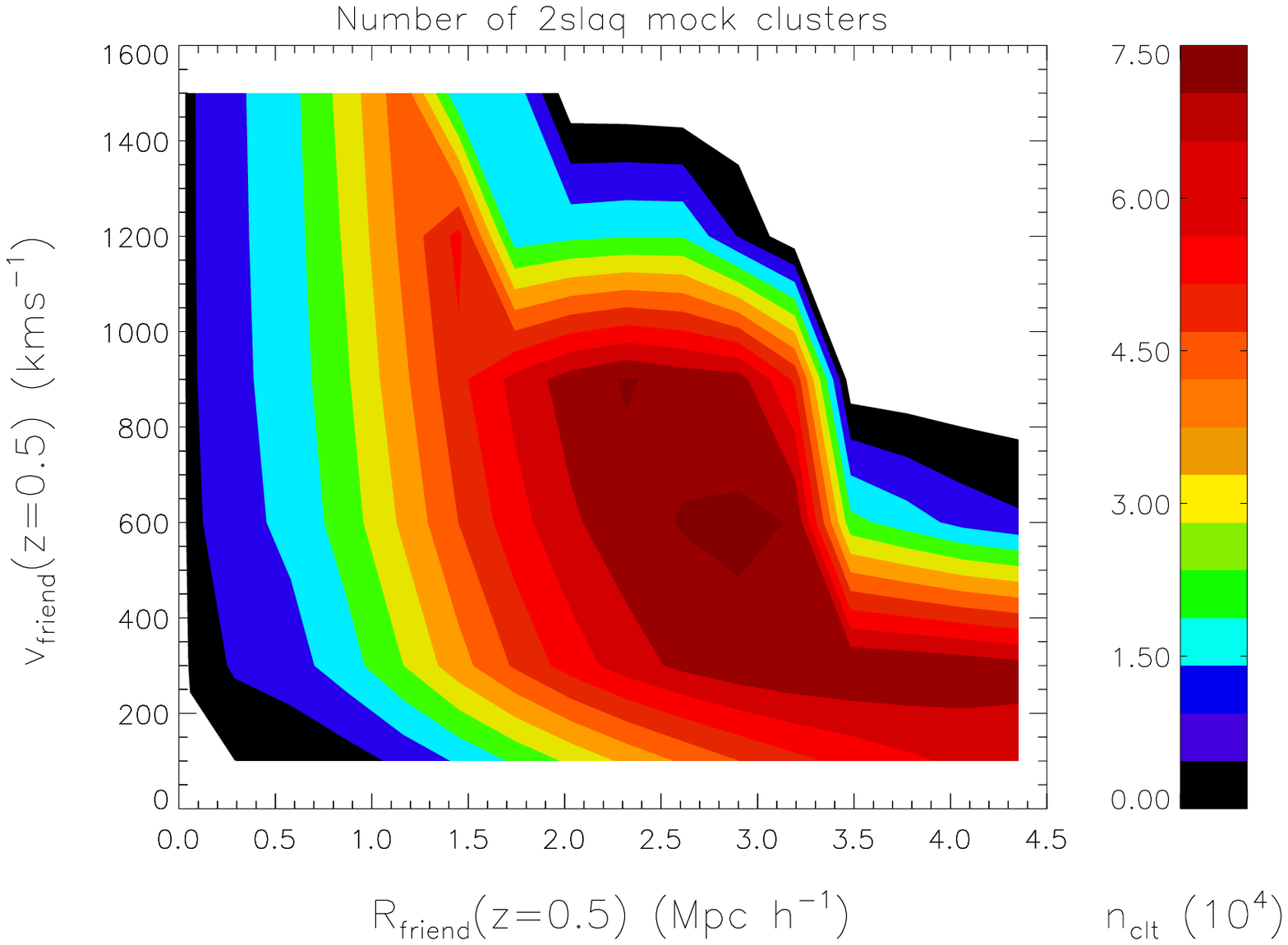}\\
	\includegraphics[width=8cm]{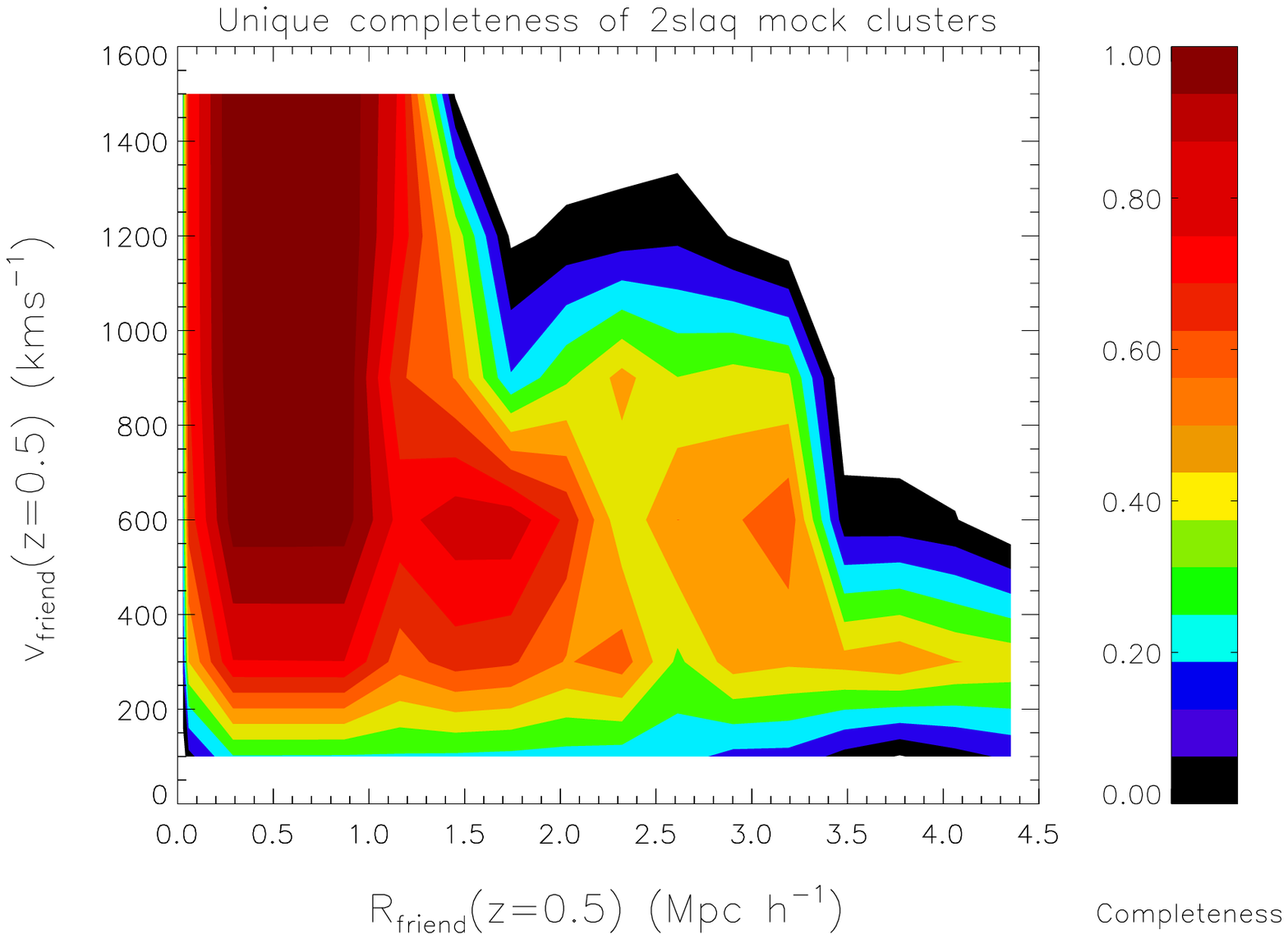}  
	\includegraphics[width=8cm]{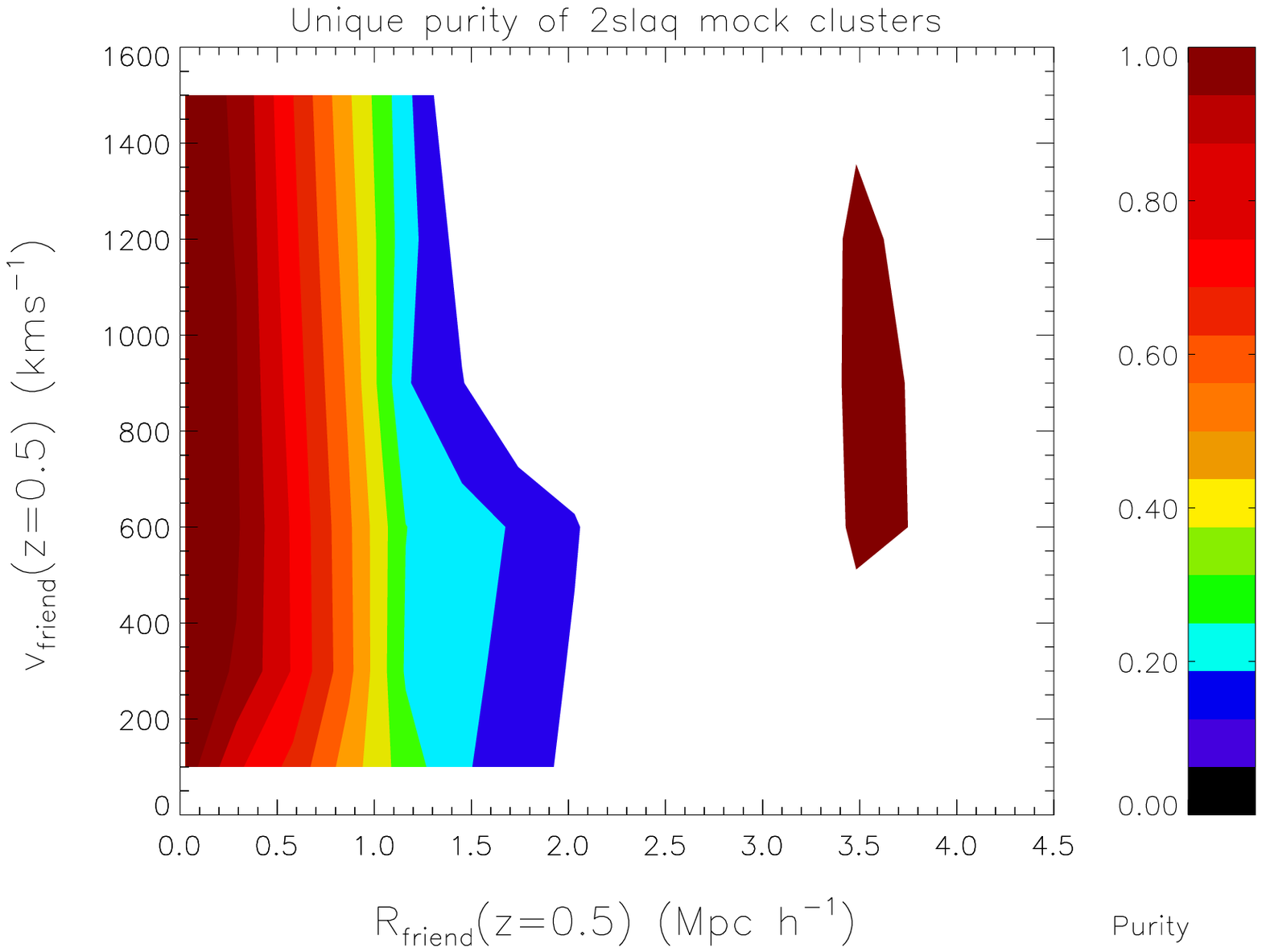}
	\includegraphics[width=8cm]{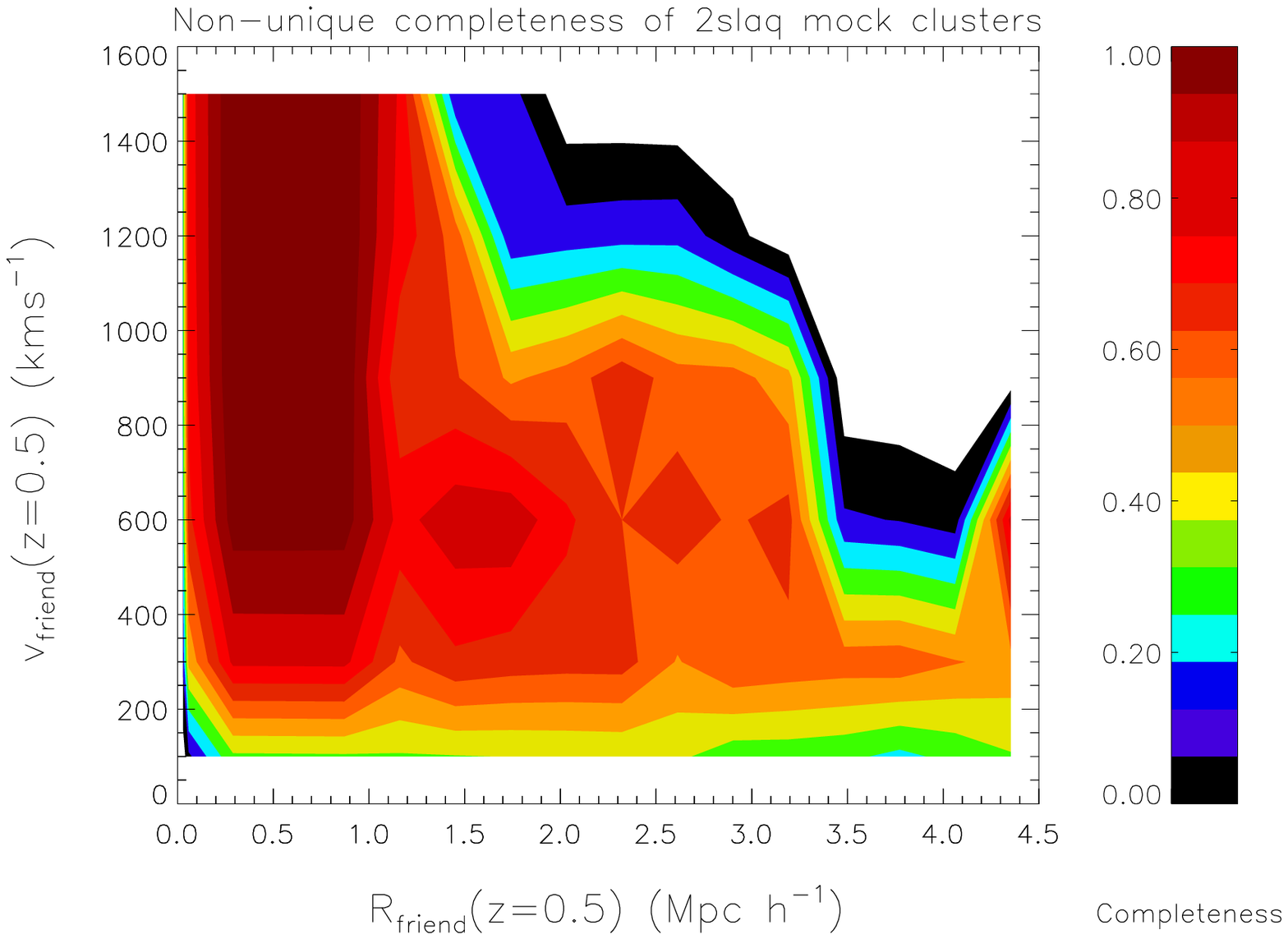}  
	\includegraphics[width=8cm]{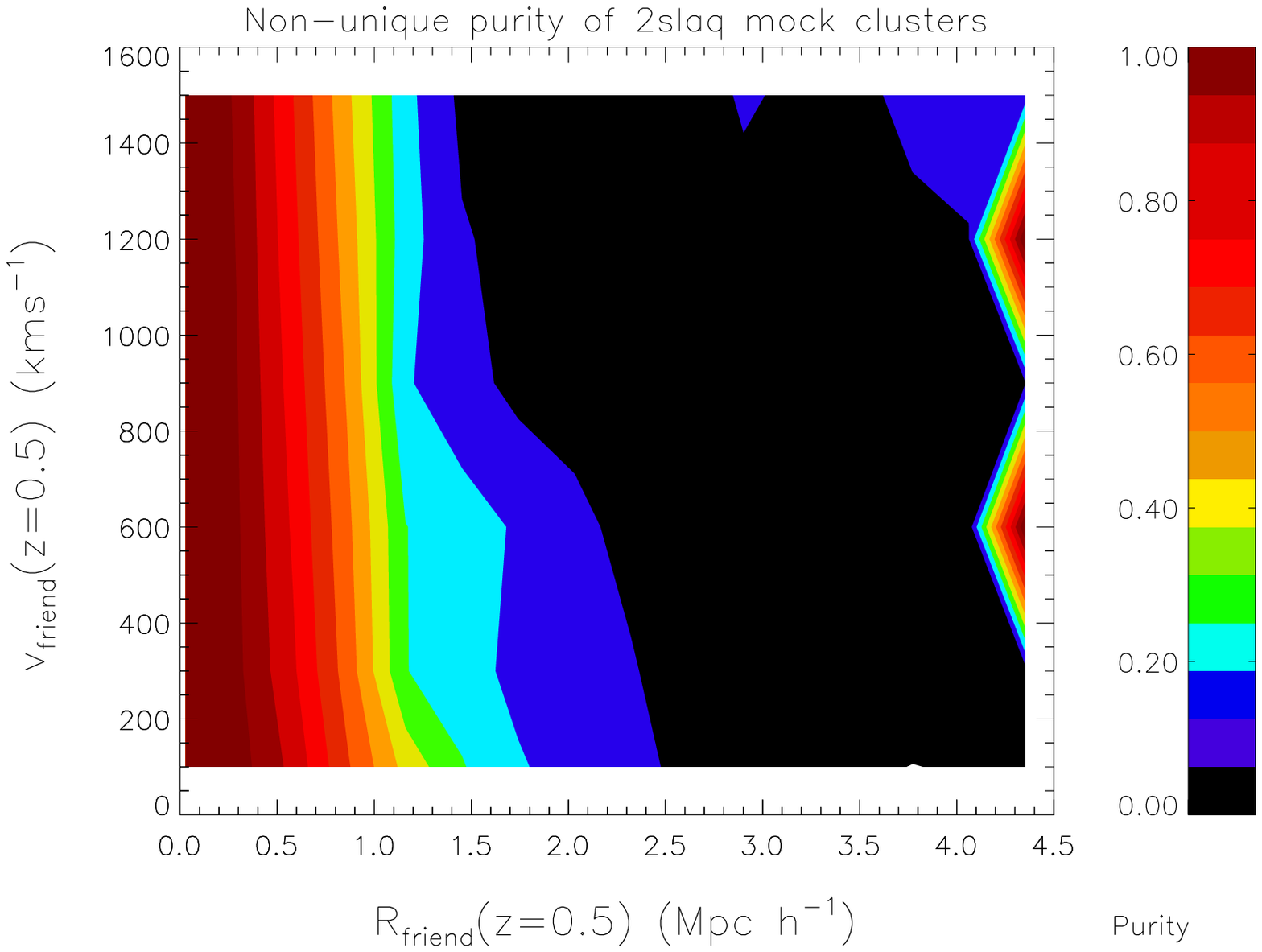}
	\caption{{Number of clusters detected in the mock halo catalogue as a function of R$_{friend}(z = 0.5)$ and  $v_{friend}(z = 0.5)$ (top panel). Unique completeness (middle left panel) and purity (middle right panel) of the DFoF clusters relative to mock haloes as a function of $R_{friend}(z = 0.5)$ and $v_{friend}(z = 0.5)$ . Non-unique completeness (bottom left panel) and purity (bottom right panel) of the DFoF clusters relative to mock haloes as a function of $R_{friend}(z = 0.5)$ and $v_{friend}(z = 0.5)$.}\label{fig:rfriends00}}
\end{figure*}
\begin{figure*}
	\centering
	\includegraphics[width=8cm]{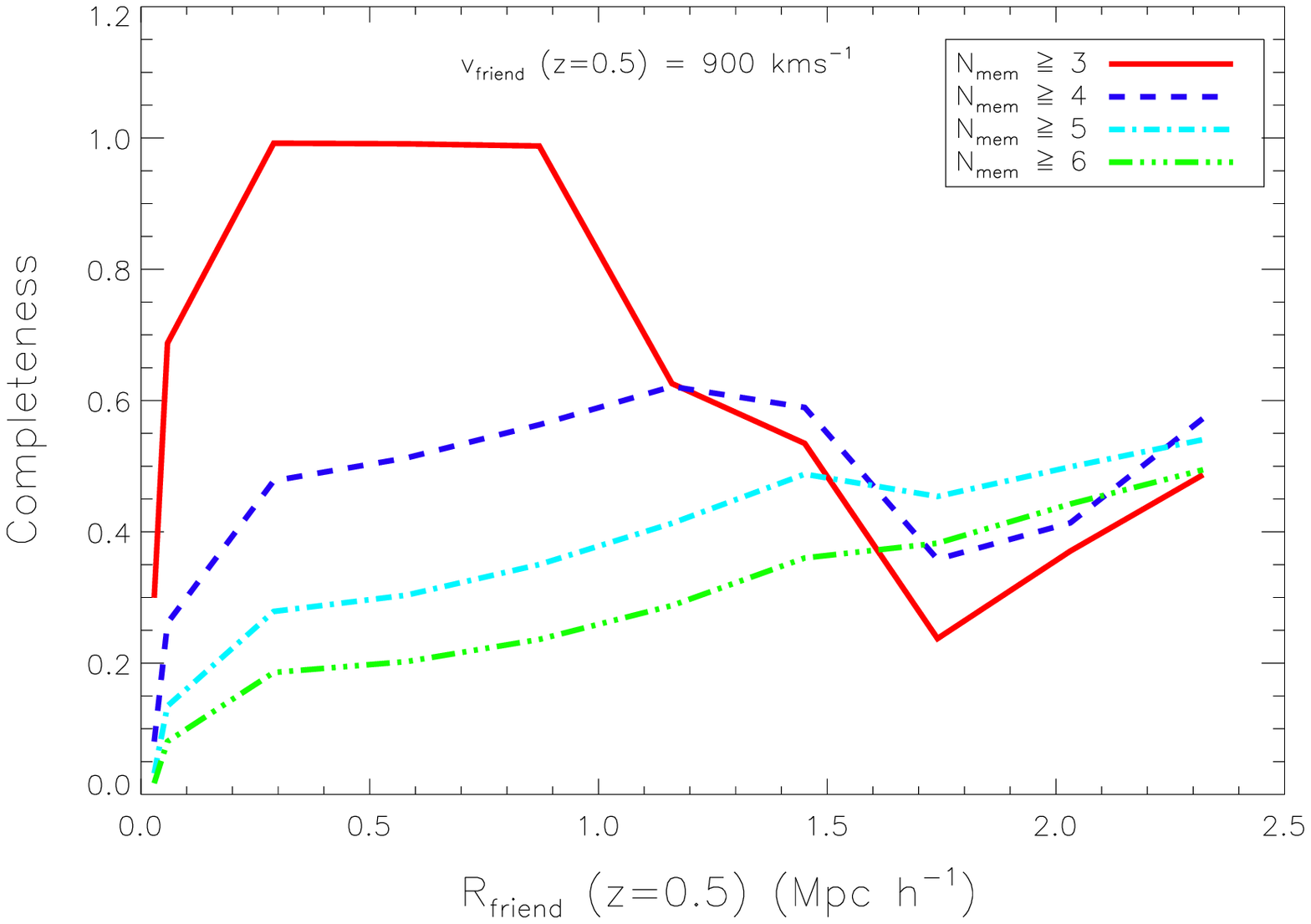} 
	\includegraphics[width=8cm]{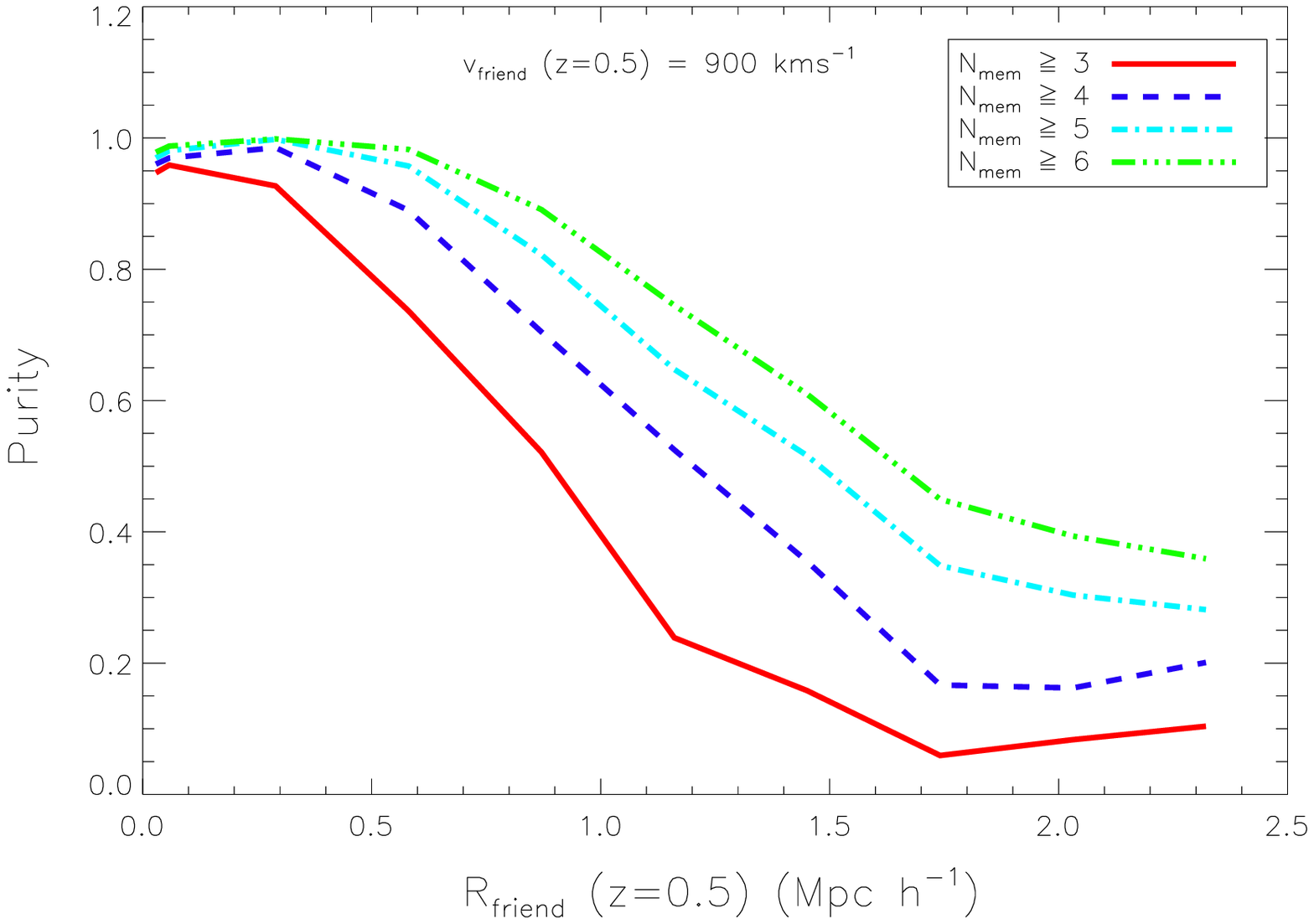}  
	\caption{{Completeness (left panel) and purity (right panel) as a function of $R_{friend}(z=0.5)$ for a fixed value of $v_{friend}(z=0.5) = 900$ kms$^{-1}$. The red solid line shows clusters with $N_{mem} \geq 3$, the blue dashed line shows clusters with $N_{mem} \geq 4$, the light blue dot-dashed line shows clusters $N_{mem} \geq 5$ and the green triple dot-dashed line shows clusters with $N_{mem} \geq 6$.}\label{fig:pure_test}}
\end{figure*}
\begin{figure*}
\includegraphics[width=8cm]{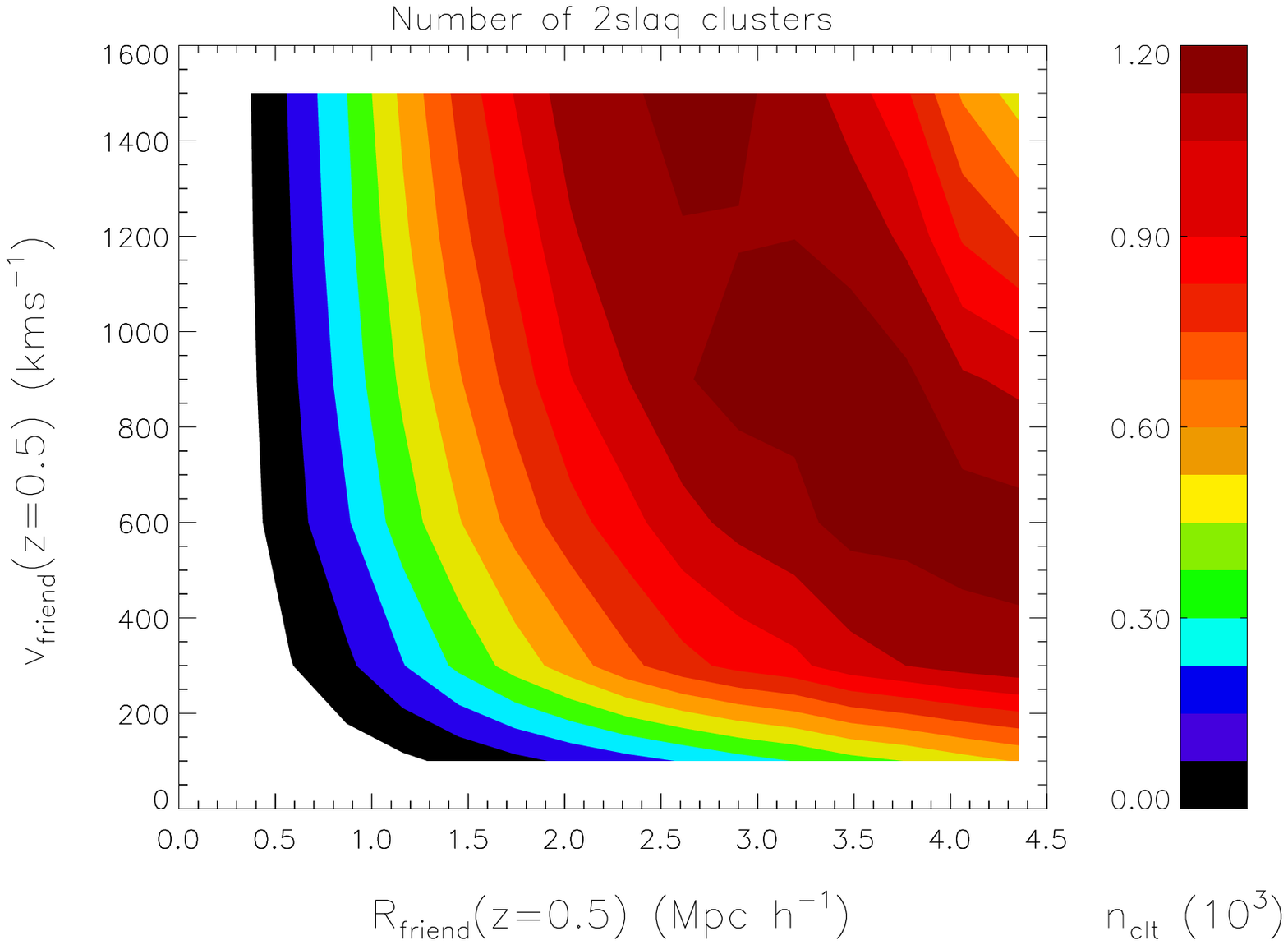} 
\caption{{Number of clusters detected in the 2SLAQ catalogue as a function of $R_{friend}(z = 0.5)$ and $v_{friend}(z = 0.5)$.}\label{fig:rfriends01}}
\end{figure*}

In order to determine the completeness and purity of the DFoF clusters relative to the simulation haloes, a membership matching code was implemented. The technique involves looking at the galaxy members assigned to each cluster by the DFoF code and matching these to the original halo member galaxies. The clusters are examined in descending order of richness to ensure that the largest clusters are the first to be matched to mock haloes. DFoF clusters may contain contributions from several mock haloes, therefore each cluster is matched to the mock halo with the highest number of shared members. Thus, if two haloes are merged into one cluster only one of the two haloes will be matched.  

For the purposes of this paper we examine two matching scenarios: a) a strict unique-matching regime, b) a less strict multiple-matching regime.  In the first case each cluster is uniquely matched to one halo and any cluster that corresponds to a halo that has already been matched, which will be of equal or lesser richness, will be ignored. In the latter case multiple clusters are allowed to match to the same halo, which increases the completeness. The completeness and purity are defined via equations \ref{eq:complete} and \ref{eq:pure}.
\begin{equation}\label{eq:complete}
\textrm{Completeness} = \frac{N_{matches}}{N_{haloes}}\,
\end{equation}
\begin{equation}\label{eq:pure}
\textrm{Purity} = \frac{N_{matches}}{N_{clusters}}\,
\end{equation}

\noindent Where $N_{matches}$ is the total number of unique or multiple matches, $N_{clusters}$ is the total number of DFoF clusters found and $N_{haloes}$ is the total number of mock haloes. $N_{haloes}$ is calculated ignoring all mock haloes with less than 3 galaxy members as this is the detection limit of the DFoF code. 

Fig.\ref{fig:rfriends00} shows the total number of clusters found in the 2SLAQ mock (top panel), unique completeness (middle left panel), unique purity (middle right panel), non-unique completeness (bottom left panel) and non-unique purity (bottom right panel) as a function of $R_{friend}(z=0.5)$ and $v_{friend}(z=0.5)$. This figure clearly shows that the DFoF results are primarily dependent on the choice of $R_{friend}(z=0.5)$. Looking at the overall contour shape in the top panel, one can see that the number of clusters found peaks around $R_{friend}(z=0.5) = $ 2.8 Mpc $h^{-1}$ and $v_{friend}(z=0.5) = $ 900 kms$^{-1}$ after which clusters are merged together creating highly unphysical structures. This sets an upper limit on both of the parameters, although we would intuitively expect the value of $R_{friend}(z=0.5)$ to be much lower. For $v_{friend}(z=0.5) > $ 400 kms$^{-1}$, the catalogue is fully complete in the range $ 0.35 < R_{friend}(z=0.5) < $ 0.87 Mpc $h^{-1}$ and fully pure out to $R_{friend}(z=0.5) = $ 0.28 Mpc $h^{-1}$ for both the unique and non-unique regimes.

Since $R_{friend}(z=0.5)$ has a greater effect on the completeness and purity, we can fix $v_{friend}(z=0.5)$ to the limit imposed by the top panel of fig.\ref{fig:rfriends00} and examine how the completeness and purity vary with just $R_{friend}(z=0.5)$. Fig.\ref{fig:pure_test} shows the variations in completeness and purity as a function of $R_{friend}(z=0.5)$ for different richness cuts using a fixed value of  $v_{friend}(z=0.5) = $ 900 kms$^{-1}$. The richness cuts show what the completeness and purity would look like if we remove groups that have fewer members than some given threshold. It should be noted that the value of $N_{haloes}$ is unchanged and therefore haloes which have fewer members than the richness cut threshold are no longer matched. This figure indicates that the richer clusters are more pure, but less complete, as one would expect. Because it is not possible to produce a catalogue that is 100\% complete and 100\% pure, it is necessary to choose one or the other, or some compromise between the two requirements. If we choose a catalogue that is $\sim$ 100\% complete, fig.\ref{fig:pure_test} shows it can be cut by richness to improve the purity.

Finally, the DFoF code was run on the real 2SLAQ catalogue to investigate the number of clusters found for a given set of linking parameters. Fig.\ref{fig:rfriends01} shows the total number of clusters found in the 2SLAQ catalogue as a function of $R_{friend}(z=0.5)$ and $v_{friend}(z=0.5)$. Both this figure and  the top panel of fig.\ref{fig:rfriends00} have the largest $N_{\textrm{clt}}$ contours roughly in the range $2.5\leq R_{friend}(z=0.5)\leq4.0$ Mpc $h^{-1}$.

Therefore in order to obtain a fully complete catalogue with high purity and the largest number of clusters possible, based on the results in fig.\ref{fig:rfriends00}, \ref{fig:pure_test} and \ref{fig:rfriends01}, values of $R_{friend}(z=0.5)=0.87$ Mpc $h^{-1}$ and $v_{friend}(z=0.5)=900$ kms$^{-1}$ were chosen to find groups and clusters in the 2SLAQ catalogue. These values correspond to a 2SLAQ mock cluster catalogue that is 98$\%$ complete and 52$\%$ pure.

\subsection{Basic Results}
\begin{figure*}
	\centering
	\includegraphics[width=8cm]{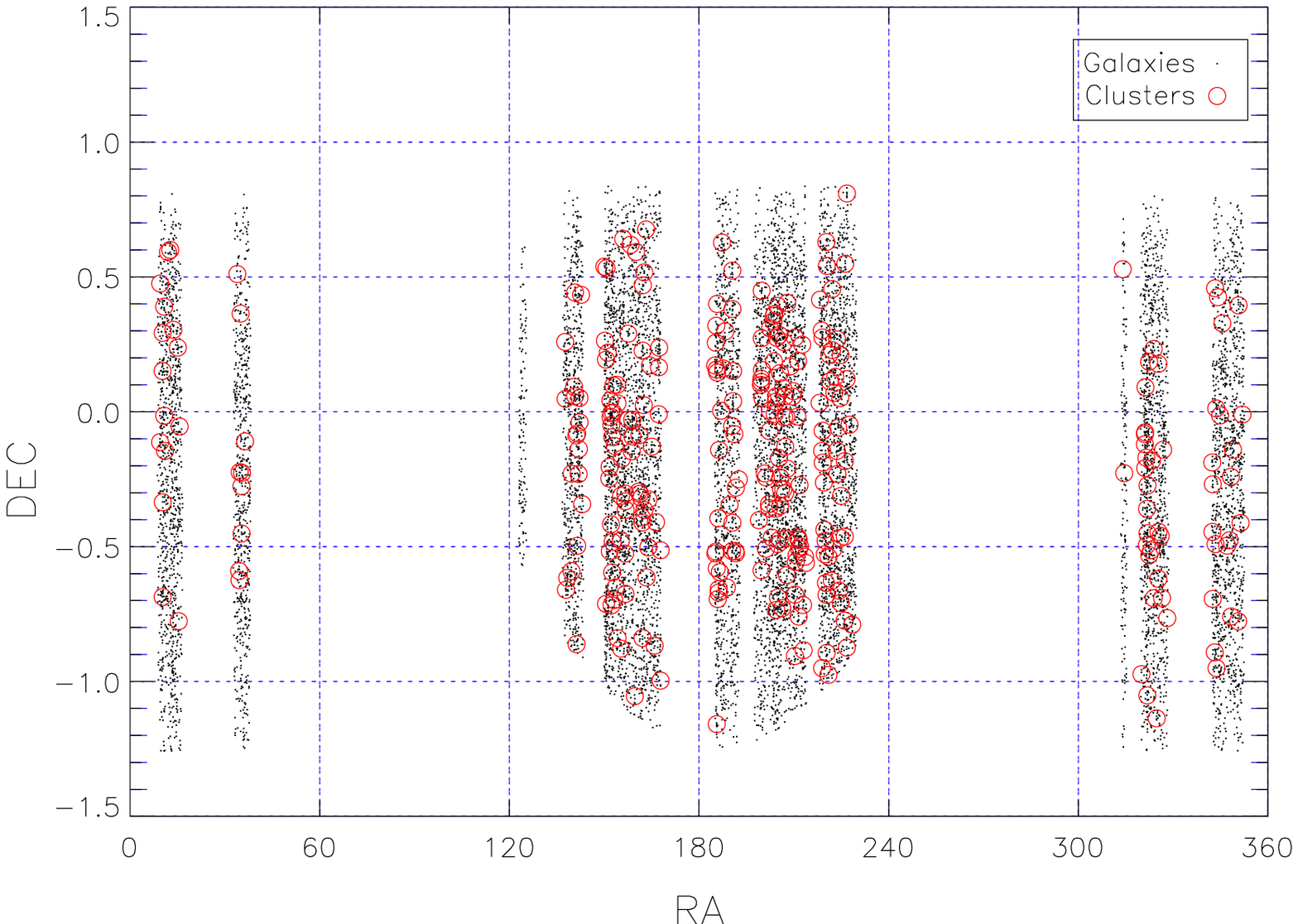}  
	\includegraphics[width=8cm]{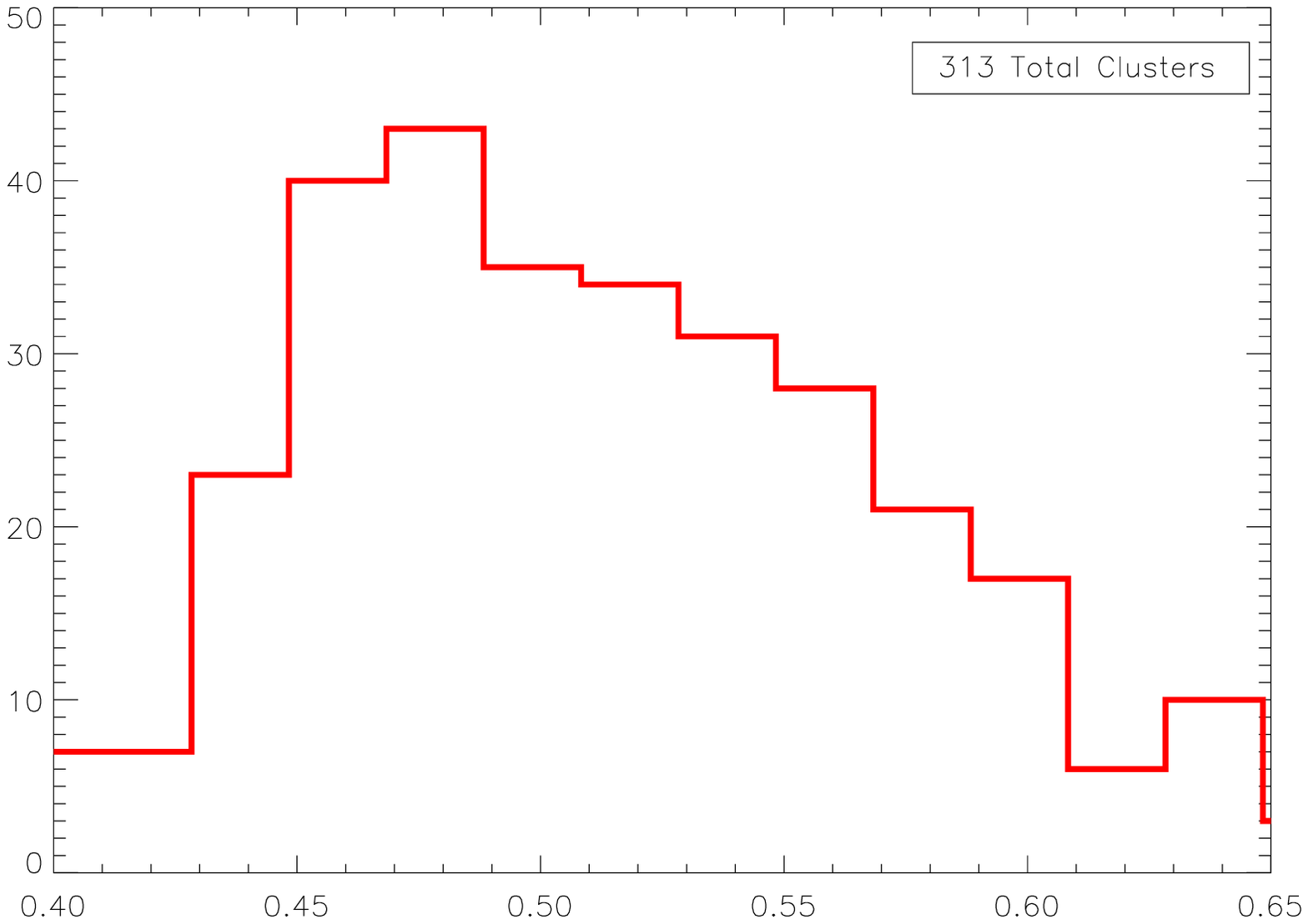}  
	\caption{{Distribution of clusters in RA and Dec (left panel). Histogram of clusters as function of redshift (right panel).}\label{fig:clusters_radec}}
\end{figure*}

Running the DFoF algorithm using a linking length of $R_{friend}(z=0.5)=0.87$ Mpc $h^{-1}$ and a velocity linking parameter of $v_{friend}(z=0.5)=900$ kms$^{-1}$ on the 13,133 galaxies in the 2SLAQ sample produced a total of 313 groups and clusters containing 1,152 member galaxies. Fig.\ref{fig:clusters_radec} shows the angular position of the centres of these groups and clusters and their distribution with respect to redshift. The cluster centres and redshifts are taken as the average values of each of the galaxy members.

\begin{figure*}
	\centering
	\includegraphics[width=8cm]{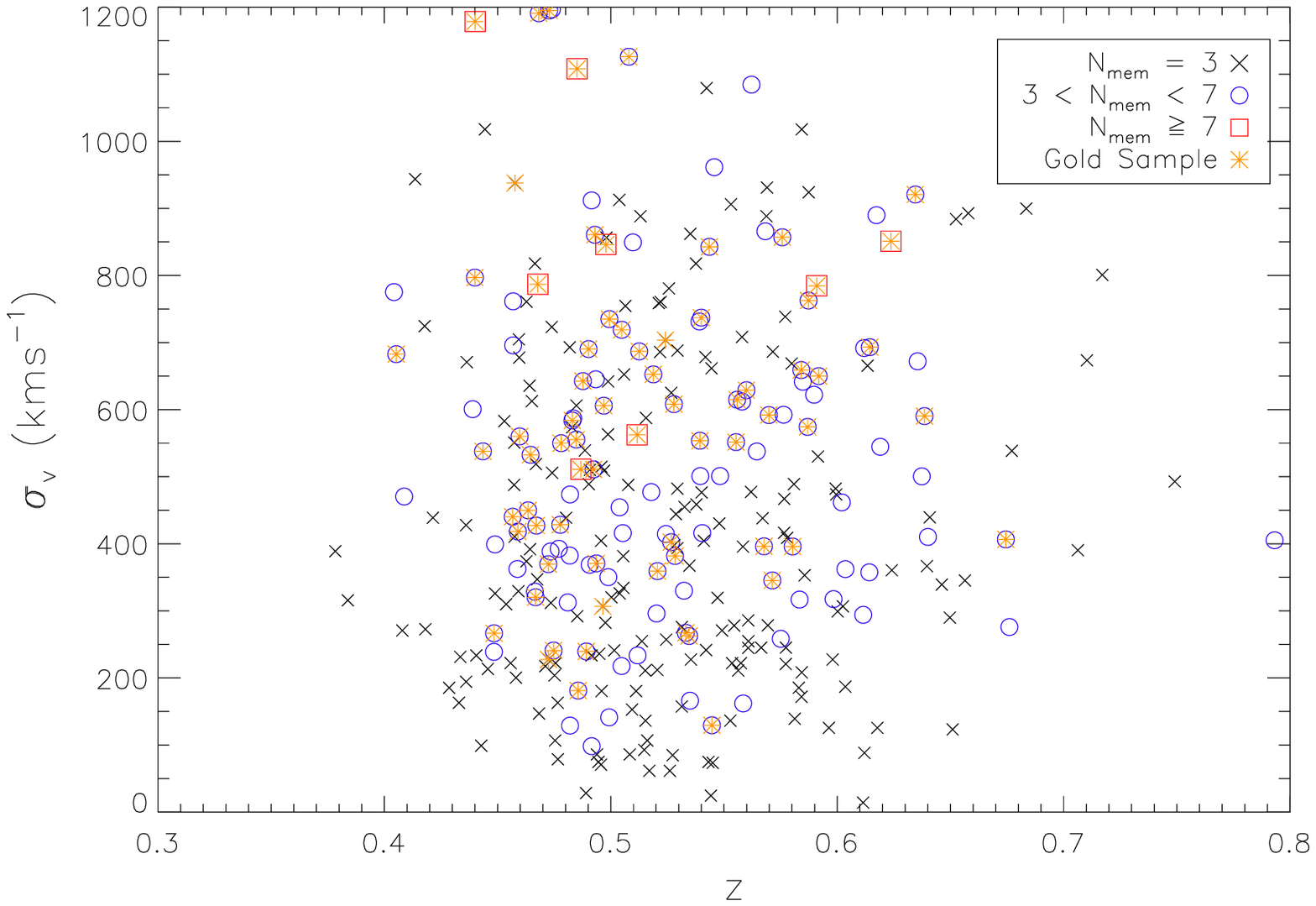}  
	\includegraphics[width=8cm]{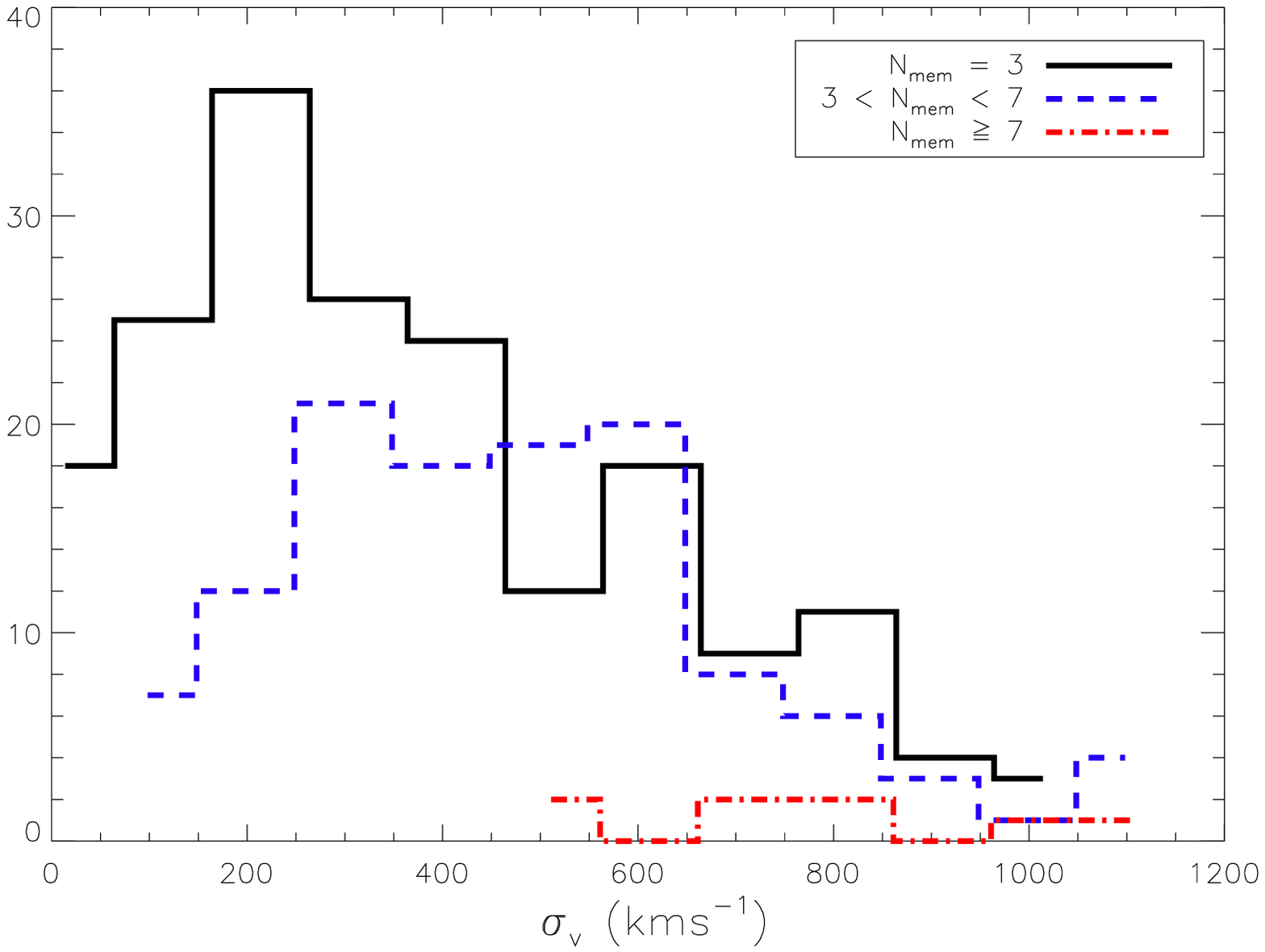}  
	\caption{{Distribution of cluster velocity dispersion, $\sigma_v$, as a function of redshift (left panel).The `x's indicate clusters with less than four members, the blue circles indicate clusters with between four and seven members and the red squares indicate clusters with more than seven members. The gold asterisks highlight the groups and clusters that form part of the gold sample (see \S 5.5). Histogram of galaxy members as a function of cluster velocity dispersion for the 3 choices of richness as in the left hand plot (right panel).}\label{fig:vel_d}}
\end{figure*}

\begin{figure*}
	\centering
	\includegraphics[width=8cm]{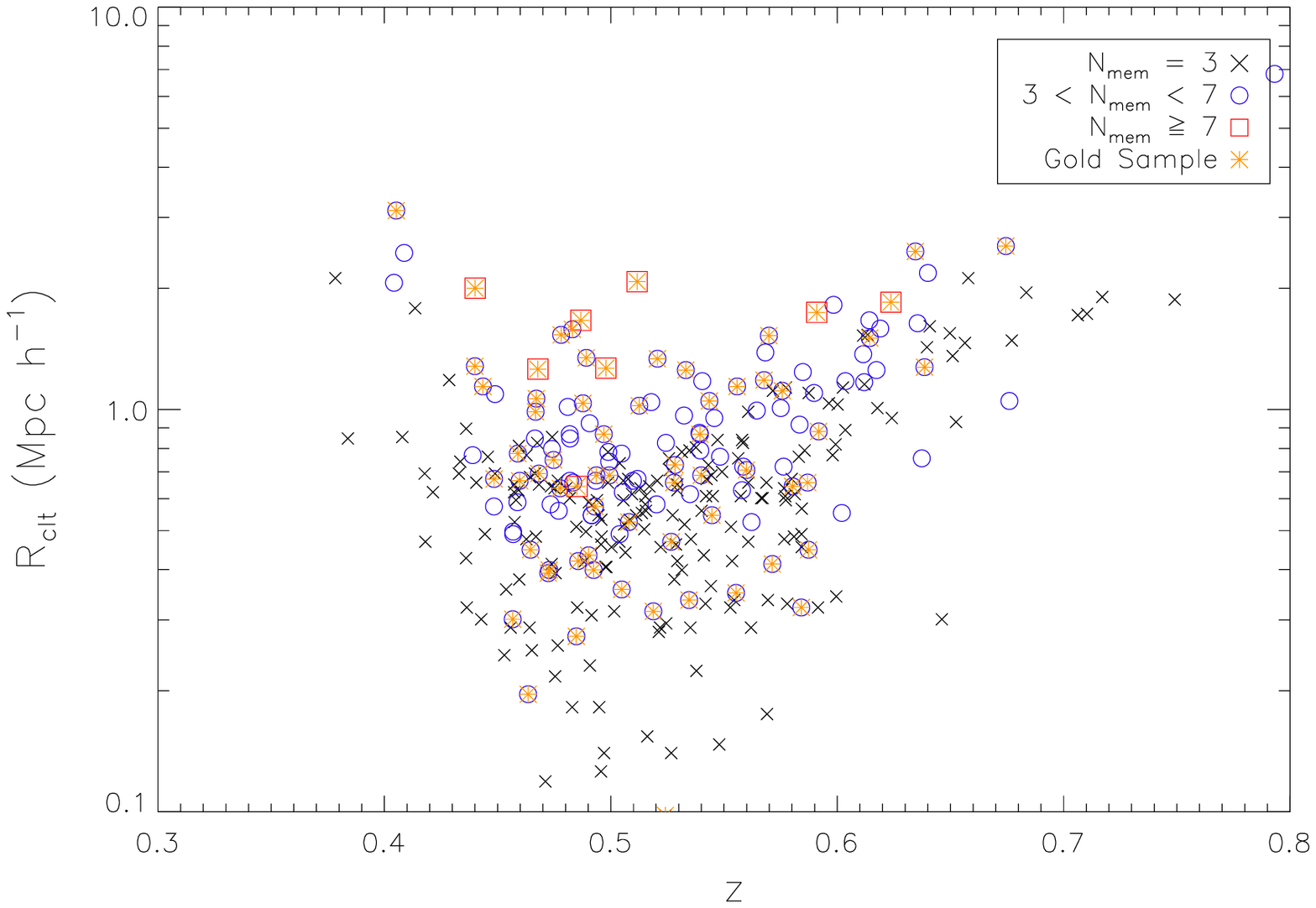}  
	\includegraphics[width=8cm]{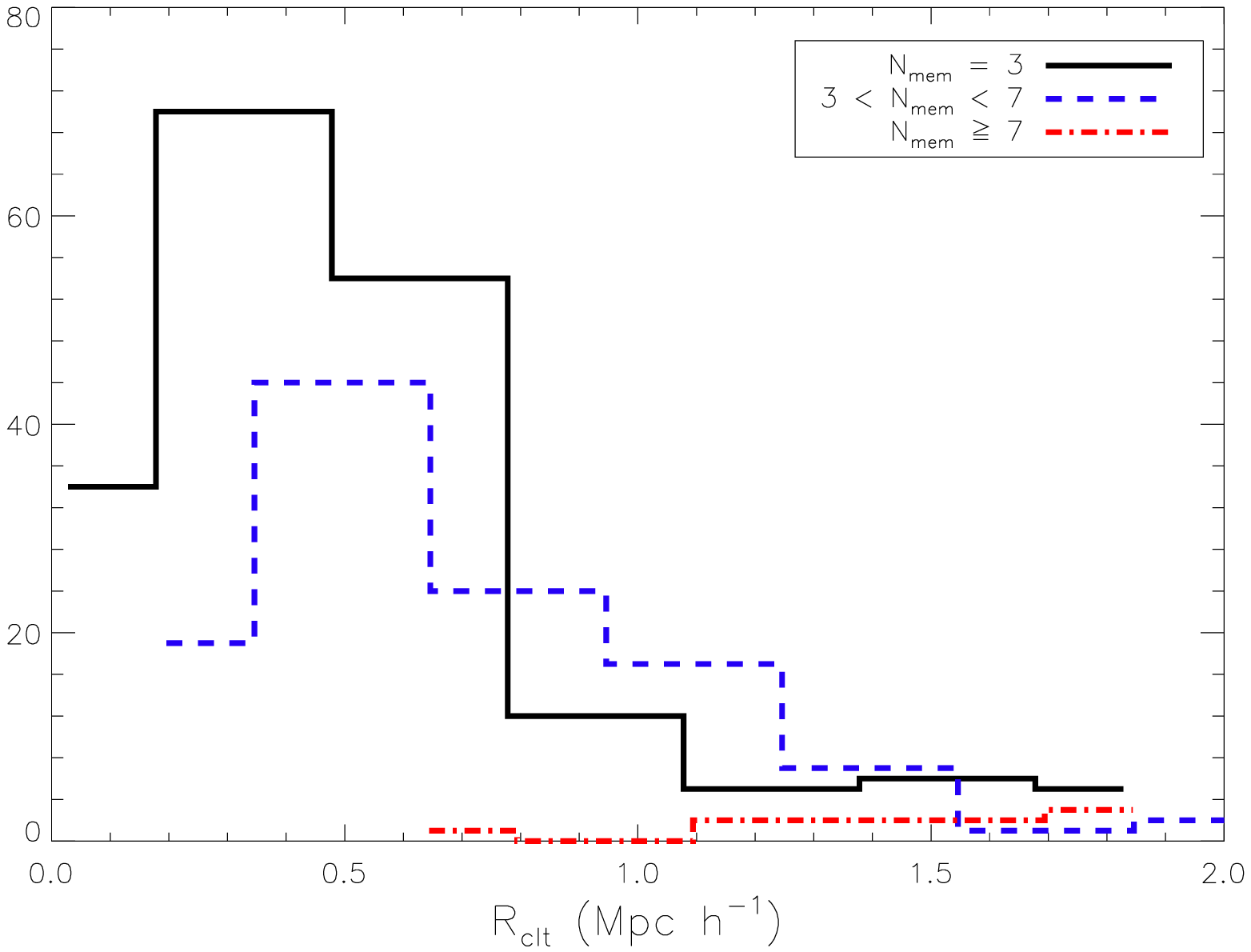}  
	\caption{{Distribution of cluster size, $R_{clt}$, as a function of redshift (left panel).The `x's indicate clusters with less than four members, the blue circles indicate clusters with between four and seven members and the red squares indicate clusters with more than seven members. The gold asterisks highlight the groups and clusters that form part of the gold sample (see \S 5.5). Histogram of galaxy members as a function of cluster size for the 3 choices of richness as in the left hand plot (right panel).}\label{fig:size_d}}
\end{figure*}

Fig.\ref{fig:vel_d} shows the distribution of cluster velocity dispersion, $\sigma_v$, as a function of redshift. Fig.\ref{fig:size_d} shows the distribution of cluster size, R$_{clt}$, as a function of redshift. Where R$_{clt}$ is defined as the projected distance from the cluster centre to the farthest galaxy member. The `x's indicate clusters with three members, the blue circles indicate clusters with between four and six members and the red squares indicate clusters with seven or more members. These plots illustrate that on average the richest clusters have the largest sizes and velocity dispersions as expected.

\subsection{Background Subtraction}
The DFoF groups and clusters were examined by looking for peaks in the photometric redshift distribution and colour-magnitude relation of SDSS galaxies around the cluster centres. Using an SQL query, regions of one square degree and centred on the FoF cluster candidates were downloaded from the SDSS DR6 catalogue. These squares were downloaded for each of the 313 DFoF groups and clusters and include all SDSS galaxies with photometric redshifts. The properties downloaded for each galaxy were: id, ra, dec, $u$-magnitude,  $g$-magnitude, $r$-magnitude, $i$-magnitude, $z$-magnitude, model $i$-magnitude, photometric redshift, photometric redshift error and star likelihood. In each of these squares all galaxies within a 1 Mpc $h^{-1}$ radius of the centre were taken as cluster members with background, regardless of redshift, and all galaxies between 3 and 7 Mpc $h^{-1}$ were used as field galaxies. Galaxies between 1 and 3 Mpc $h^{-1}$ were ignored. These limits were chosen because, on average, we would not expect galaxies father than 3 Mpc $h^{-1}$ from a real cluster centre to be genuine members. Using the DFoF cluster redshifts, the data from the corresponding SDSS squares were stacked in redshift slices of $\Delta z=0.05$ in the range $0.4\leq z\leq0.7$. The total number of groups and clusters with three or more and four or more members in each redshift slice are listed in Table-\ref{tab:zrange}.
\begin{table}
\caption{Redshift Slices}
\begin{tabular}{|c|c|cl}
	\hline
z Range & N$_{clt}$ with N$_{mem}\geq3$ &  N$_{clt}$ with N$_{mem}\geq4$  \\
	\hline
$0.40\leq z\leq0.45$ & 26  & 10 \\
$0.45\leq z\leq0.50$ & 99  & 46 \\
$0.50\leq z\leq0.55$ & 85 & 30 \\
$0.55\leq z\leq0.60$ & 62 & 24 \\
$0.60\leq z\leq0.65$ & 26 & 14\\
$0.65\leq z\leq0.70$ & 8  & 2\\
	\hline
\end{tabular}
\label{tab:zrange}
\end{table}

\begin{figure*}
	\centering
	\includegraphics[width=8cm]{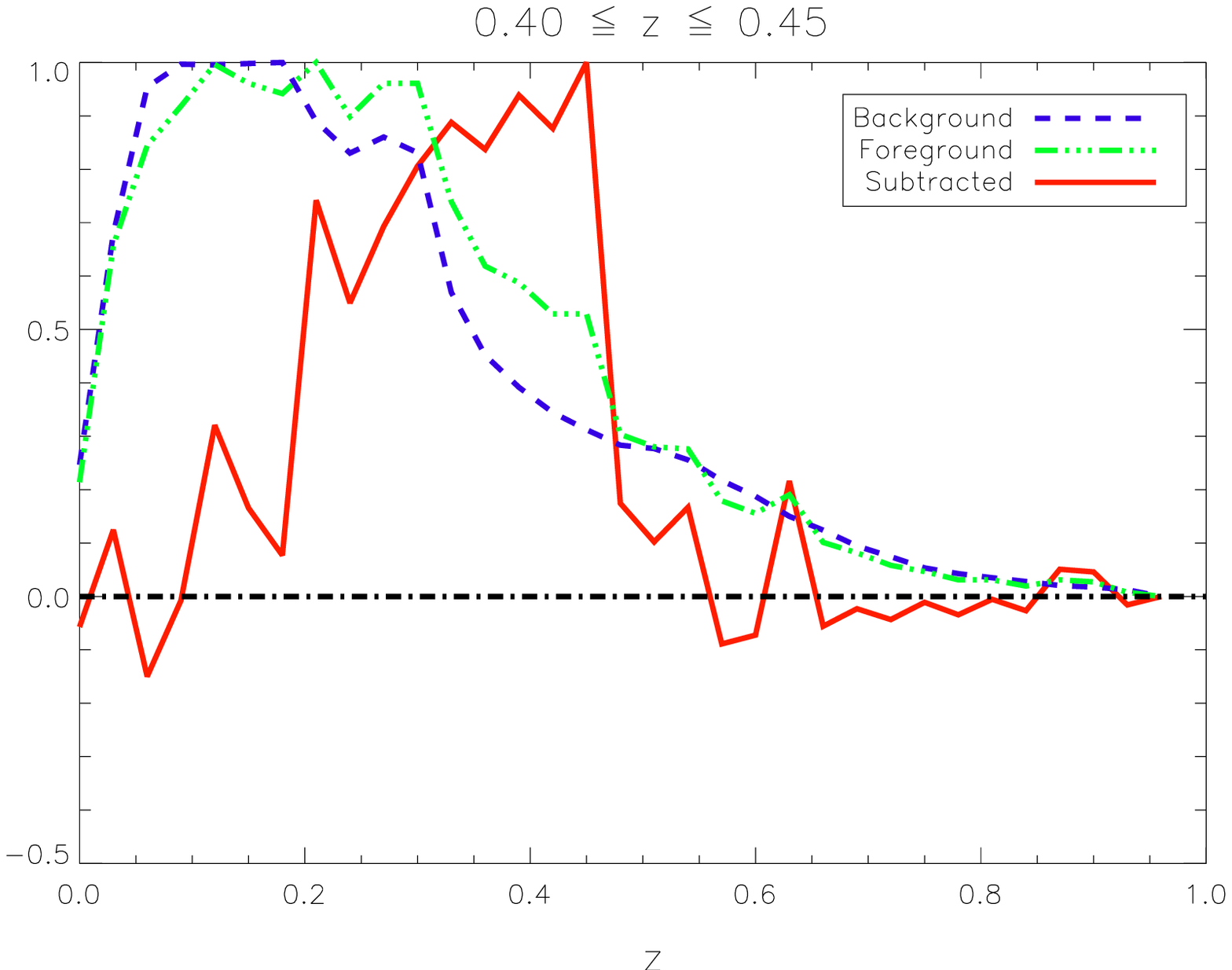} 
	\includegraphics[width=8cm]{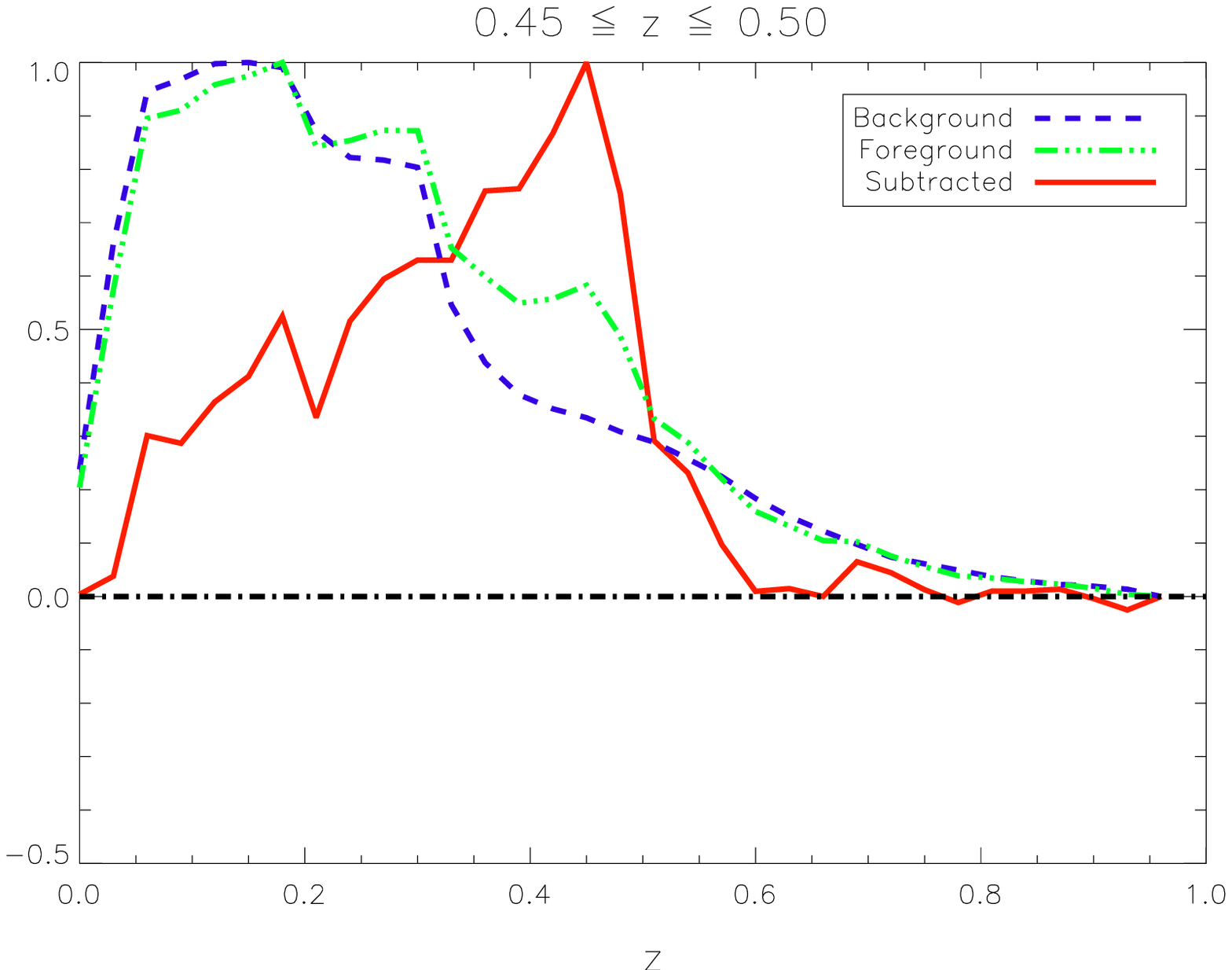} 
	\includegraphics[width=8cm]{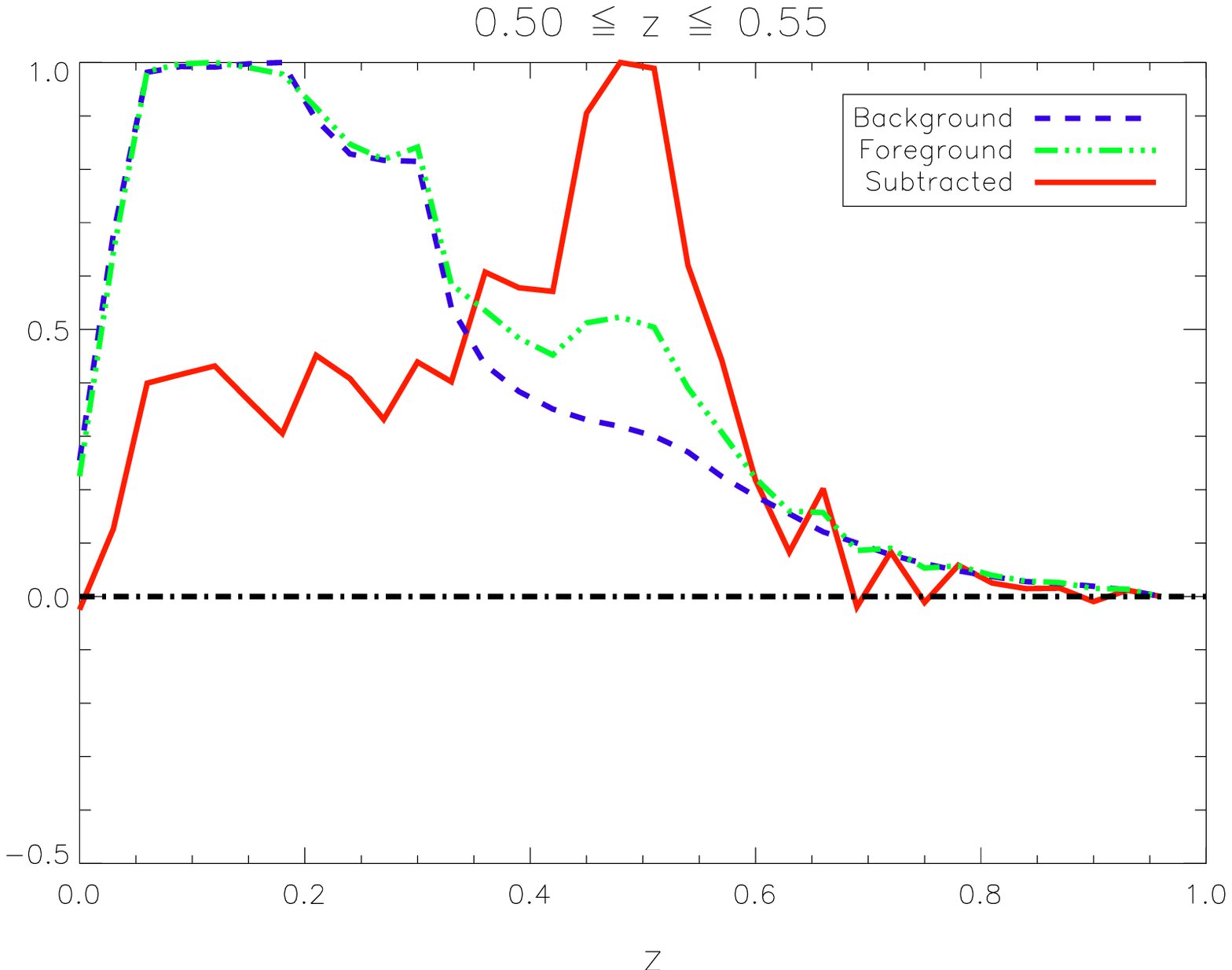} 
	\includegraphics[width=8cm]{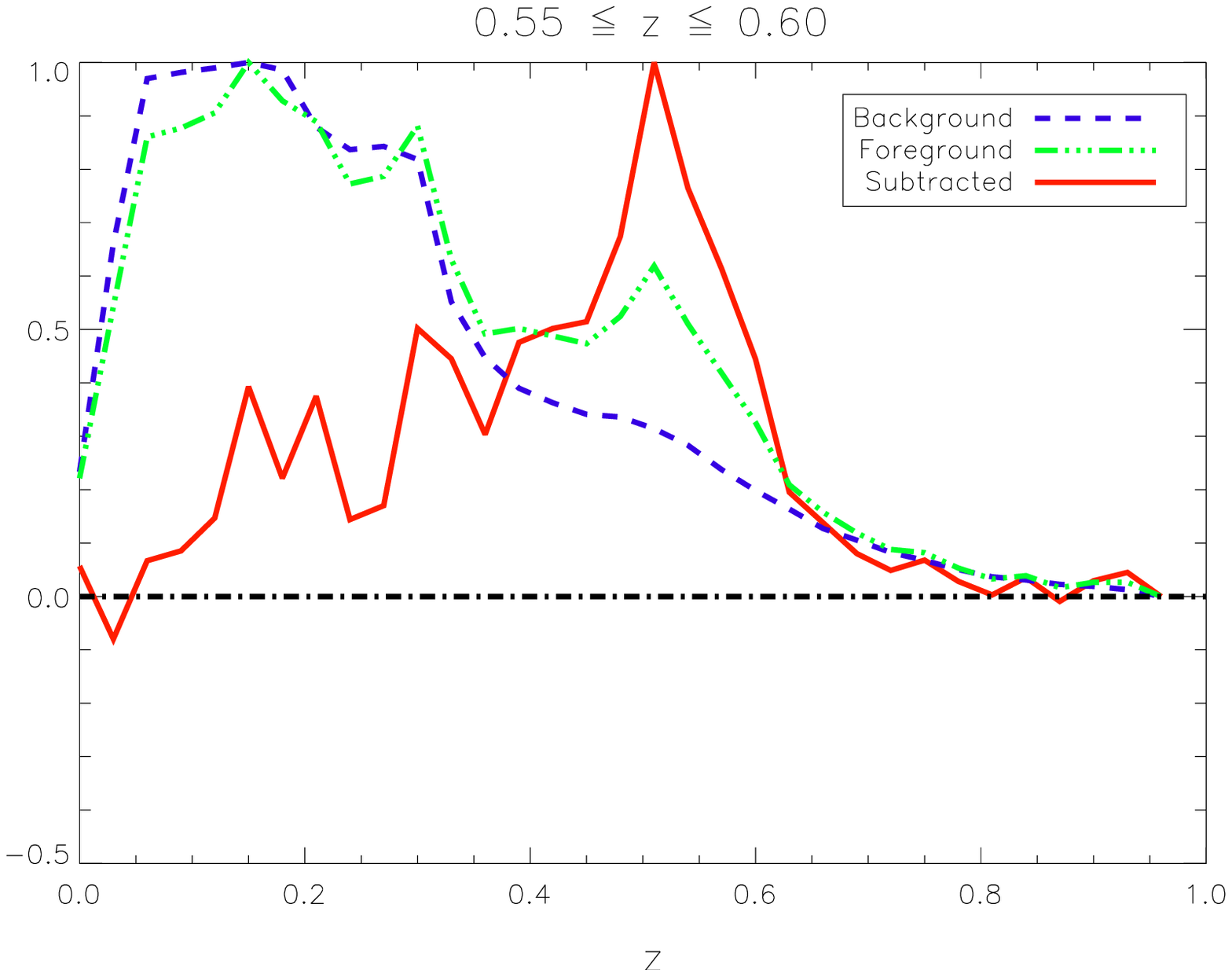} 
	\includegraphics[width=8cm]{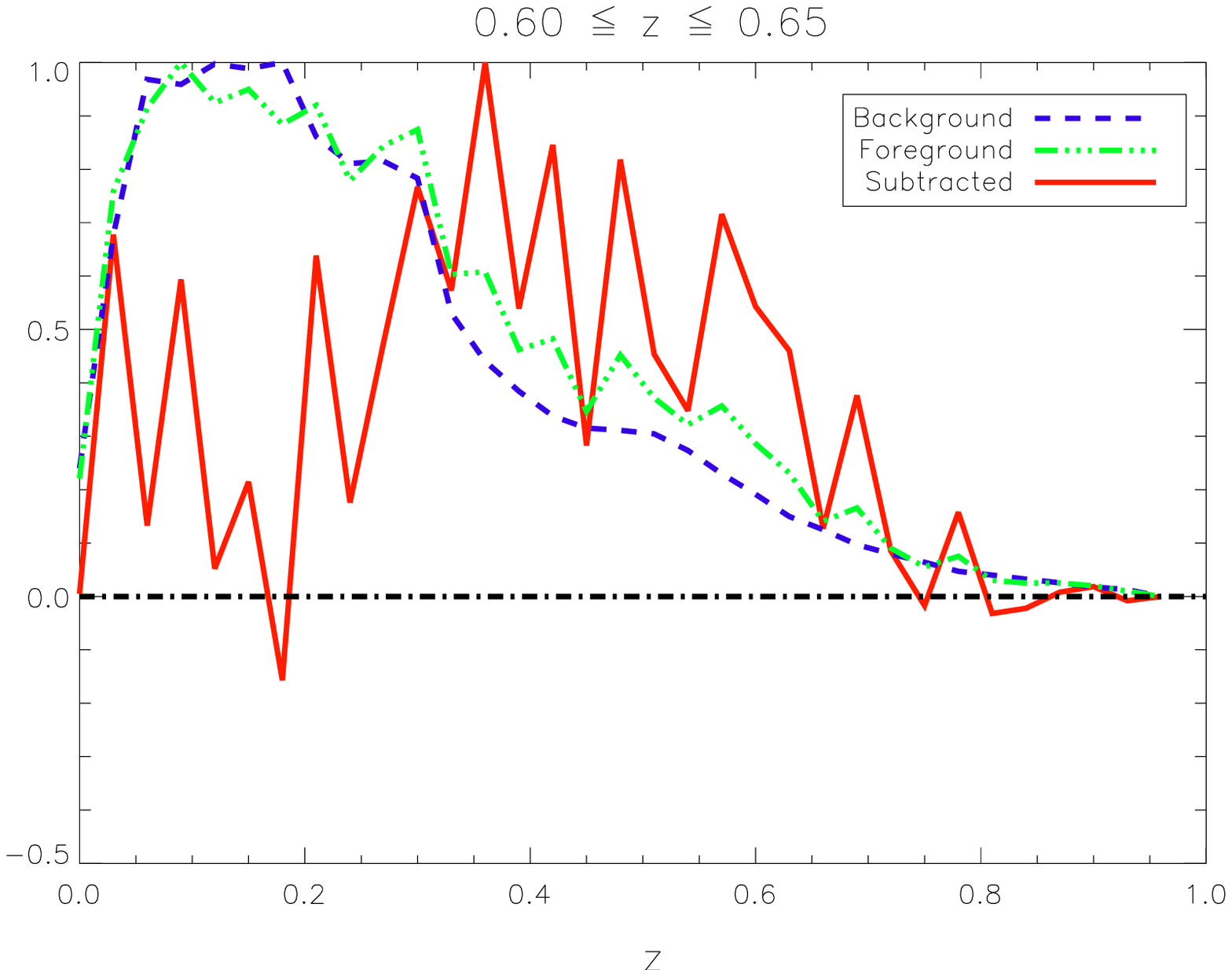} 
	\includegraphics[width=8cm]{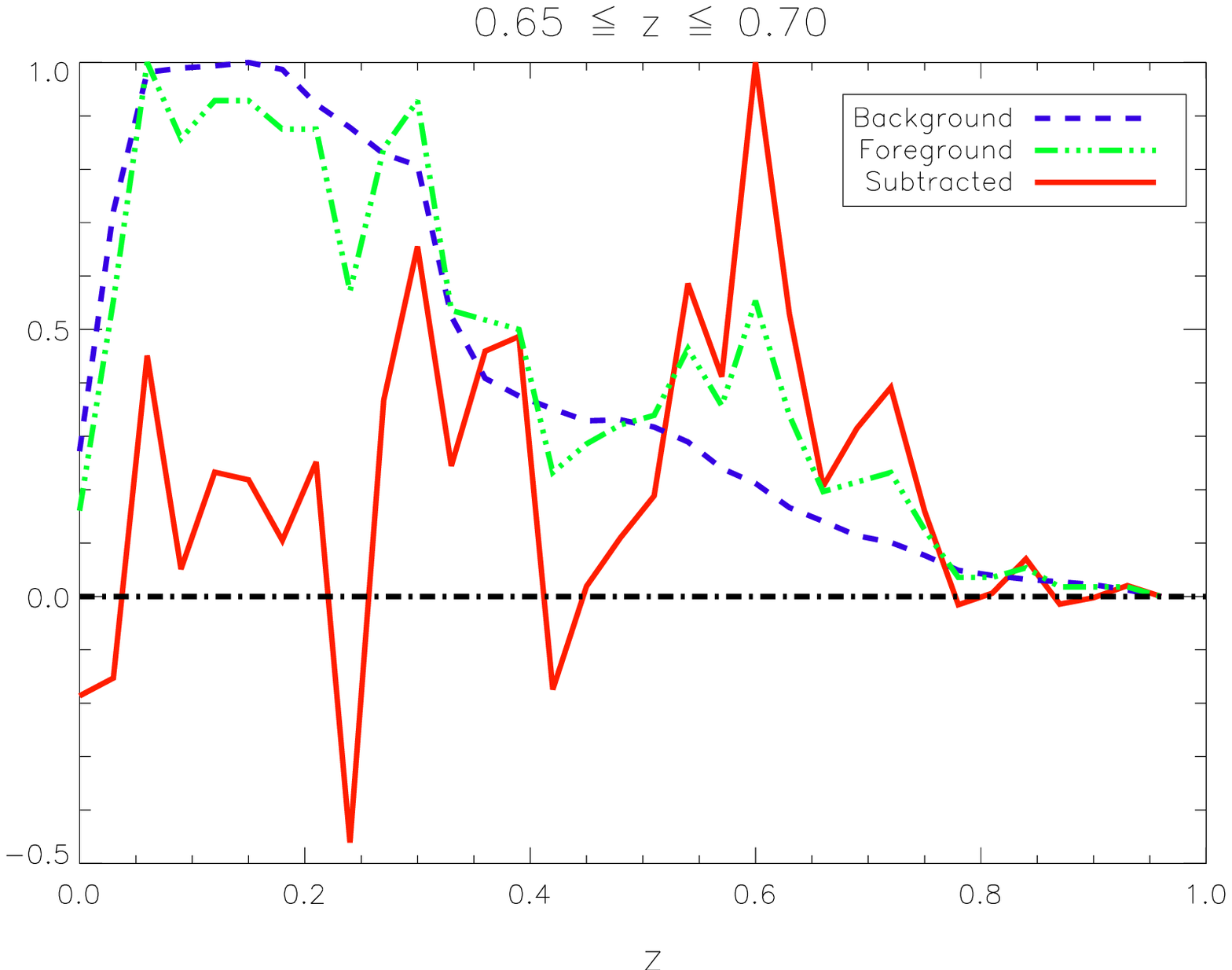}
	\caption{{Number of SDSS background subtracted galaxies as a function of photometric redshift for all groups and clusters. The blue dashed line are the background galaxies, the green dot-dashed line are the foreground galaxies and the red solid line are the background subtracted galaxies.}}\label{fig:nz_plots}
\end{figure*}

In each cluster redshift bin the stacked SDSS galaxies were binned by photometric redshift. The number of SDSS galaxies assigned to the field was then subtracted from the number of SDSS galaxies assigned as cluster members plus background in each redshift bin. This subtraction was done taking into account the difference in area between the two regions. Fig.\ref{fig:nz_plots} shows the number of SDSS background subtracted galaxies as a function of photometric redshift for all groups and clusters. The blue dashed line are the background galaxies, the green dot-dashed line are the foreground galaxies and the red solid line are the background subtracted galaxies. In these plots the background subtraction should average out to zero if there is no overdensity in the field. A sharp peak can clearly been seen in each of the plots at the cluster redshift range. This trend is reliable up to $z\sim 0.6$ after which the peaks cannot clearly be distinguished. This a strong indication that the DFoF cluster candidates are genuine structures. The discrepancies seen after $z\sim 0.6$ are due to the relatively low number of clusters at these redshifts plus a result of poor photometric redshift estimates for more distant objects in the SDSS catalogue. 

\begin{figure*}
	\centering
	\includegraphics[width=8cm]{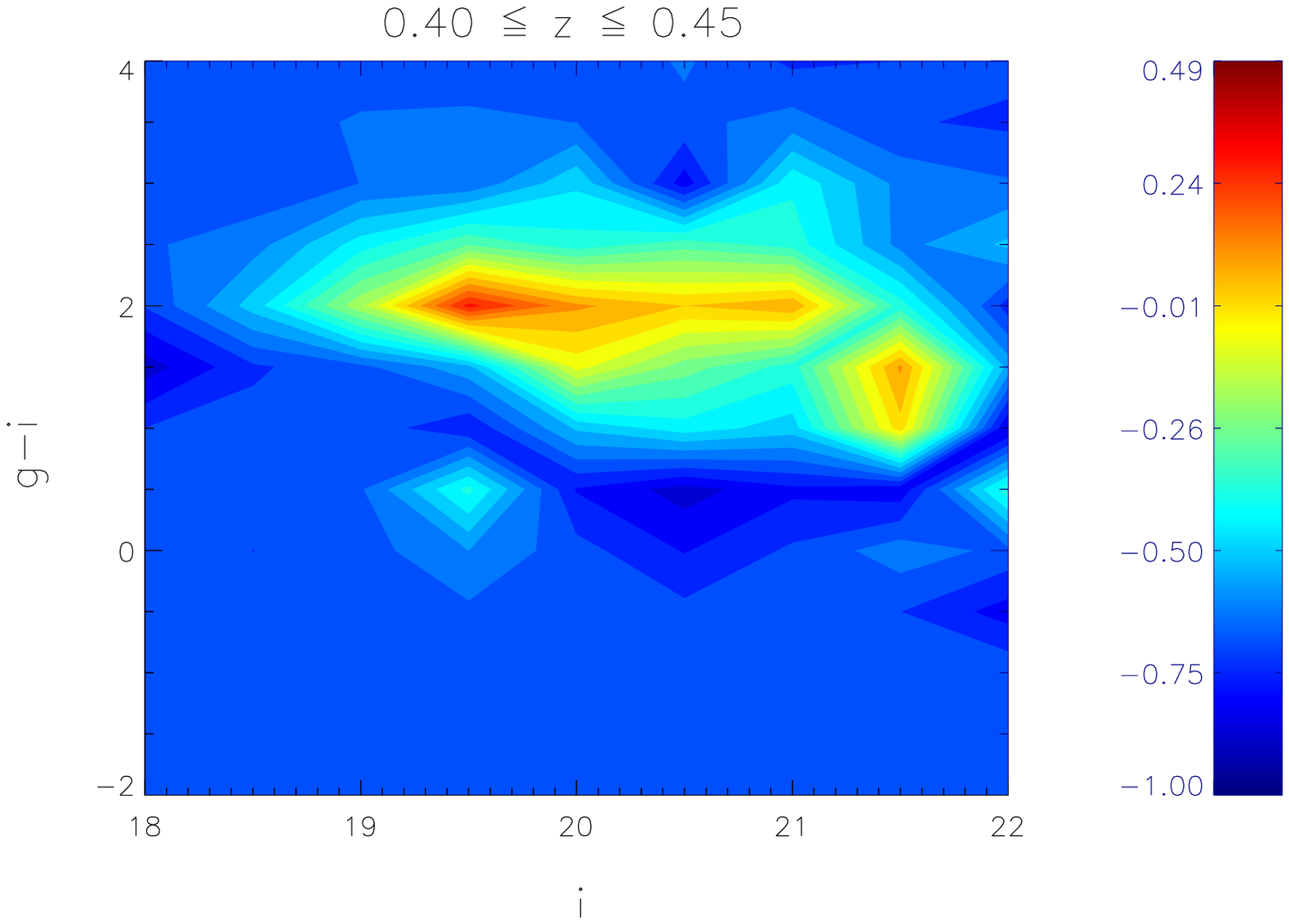} 
	\includegraphics[width=8cm]{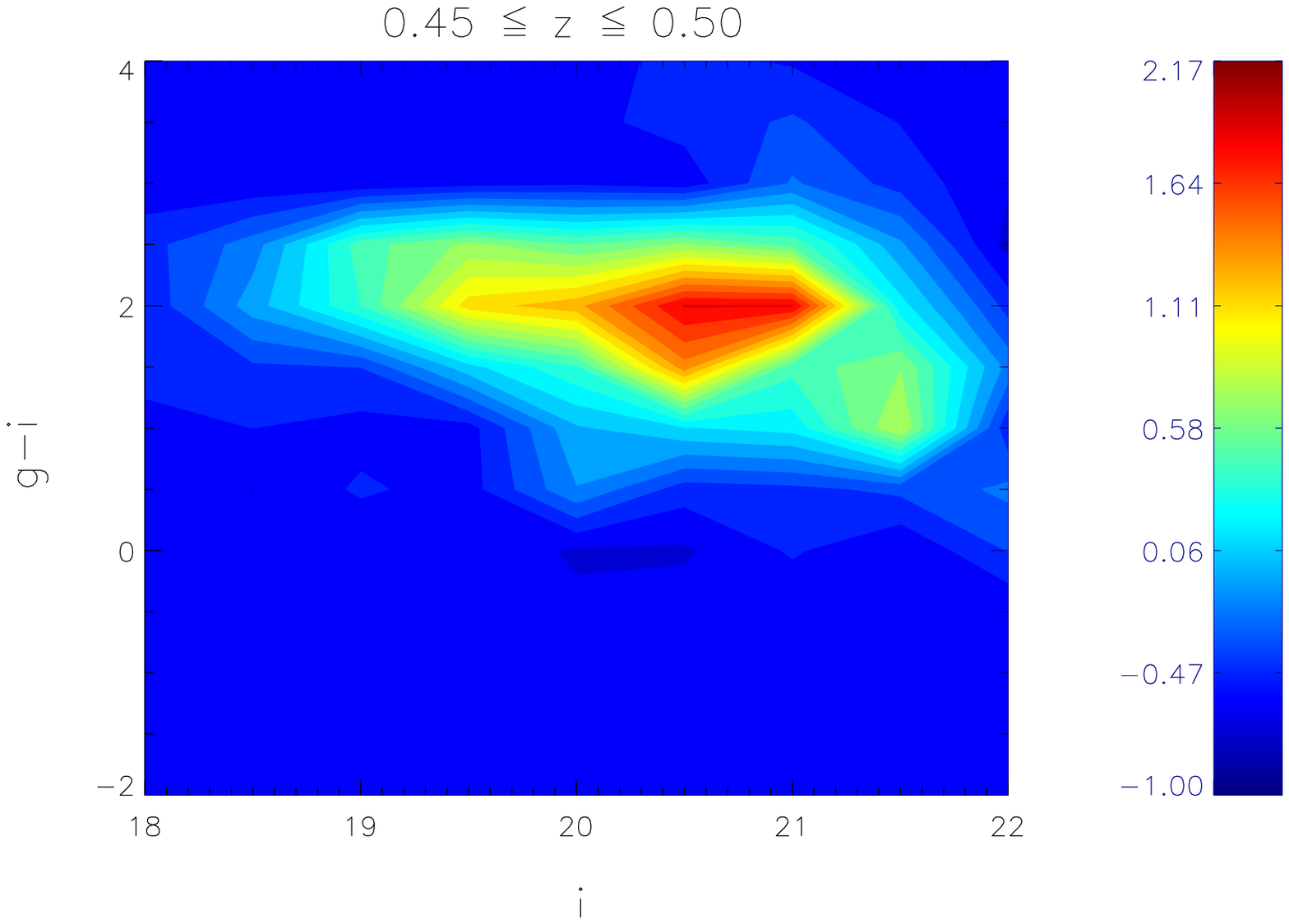} 
	\includegraphics[width=8cm]{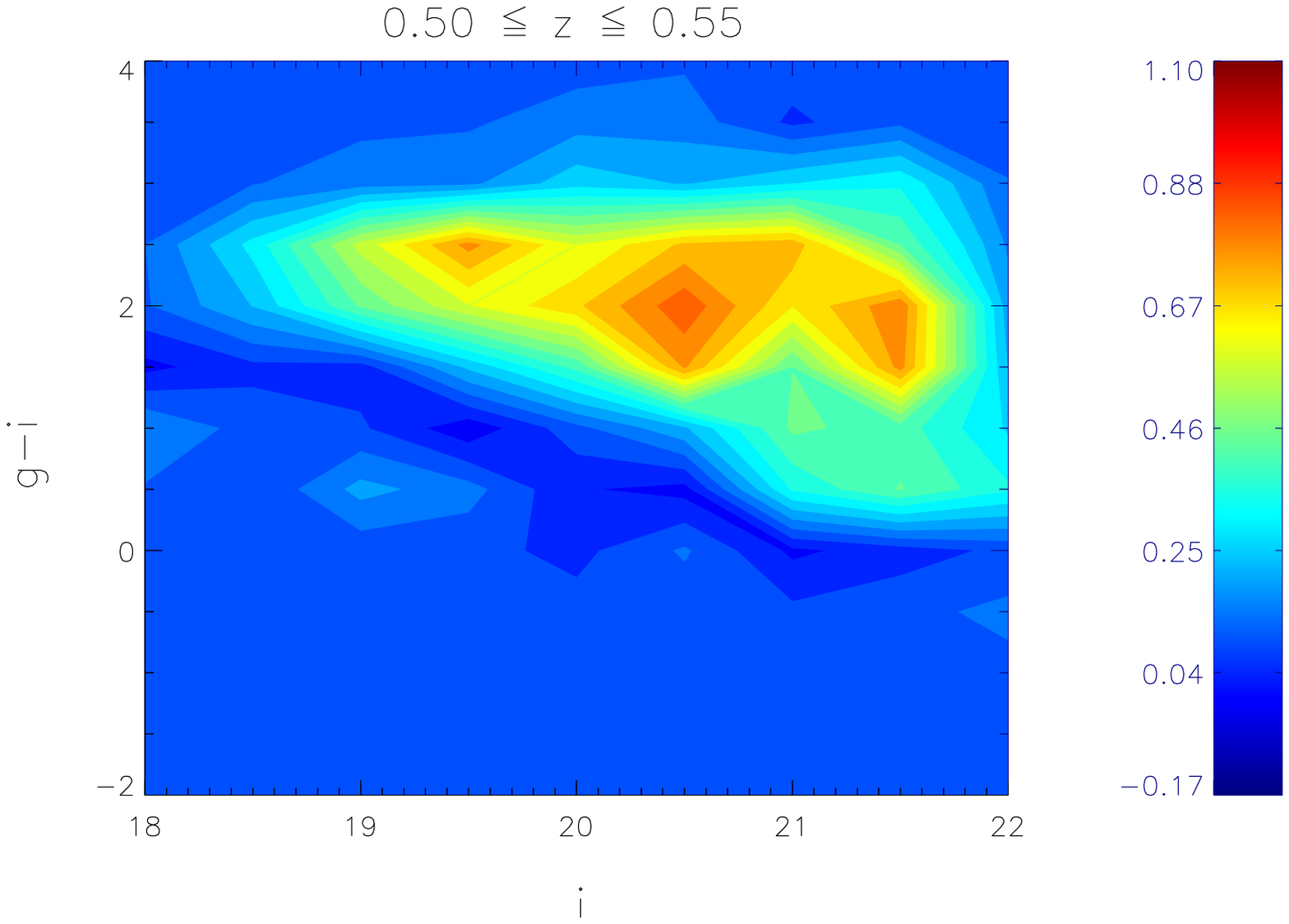} 
	\includegraphics[width=8cm]{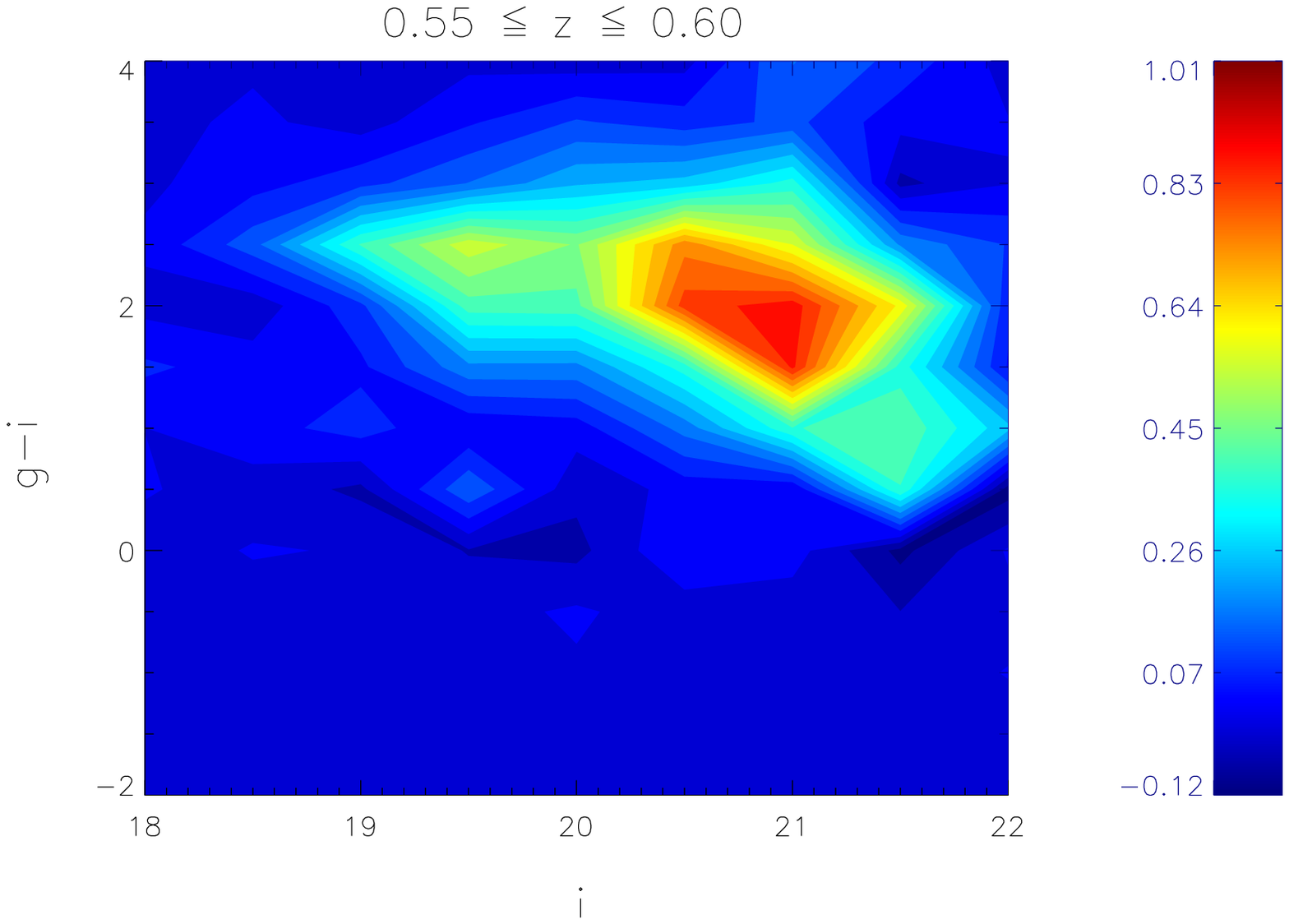} 
	\includegraphics[width=8cm]{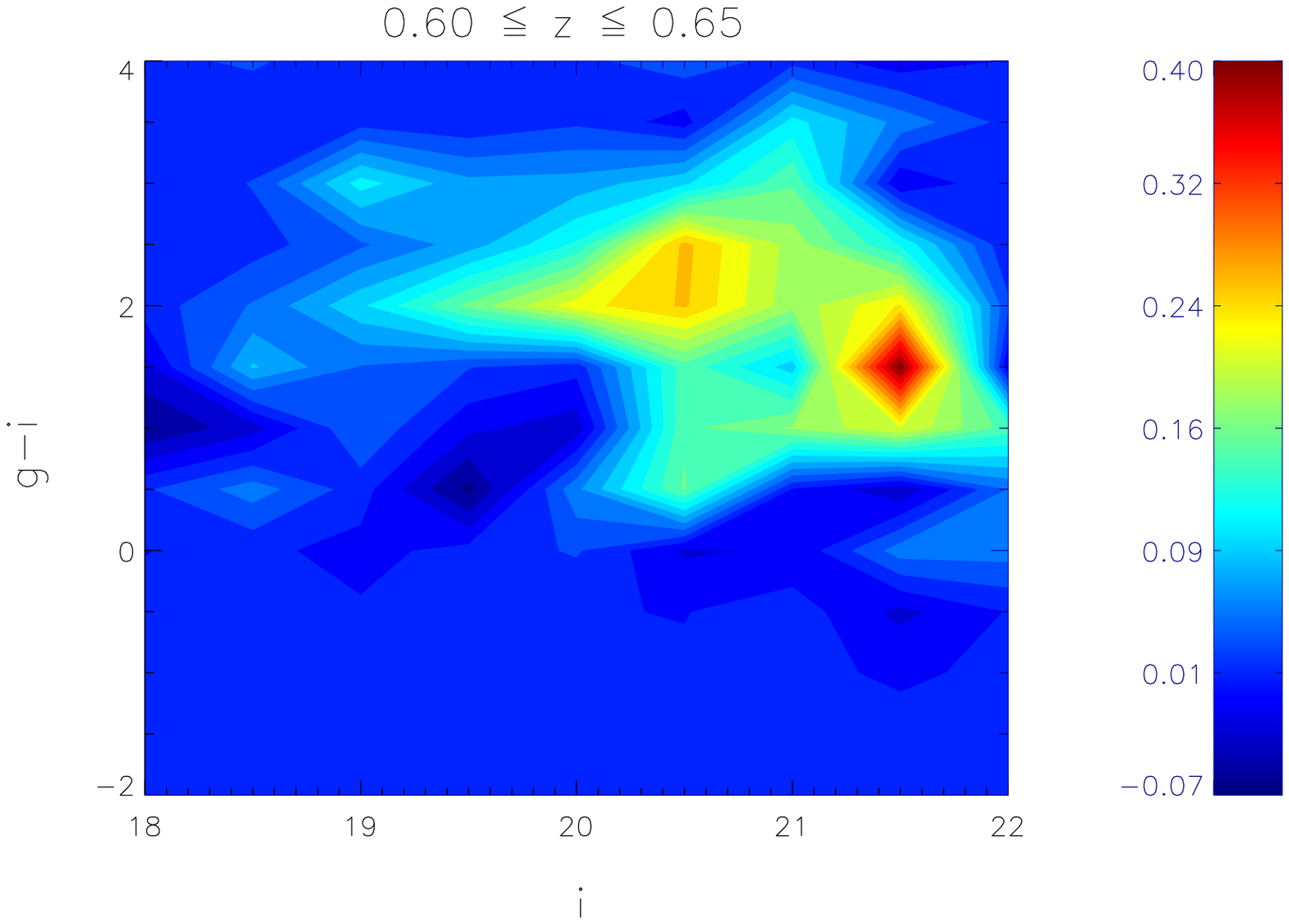} 
	\includegraphics[width=8cm]{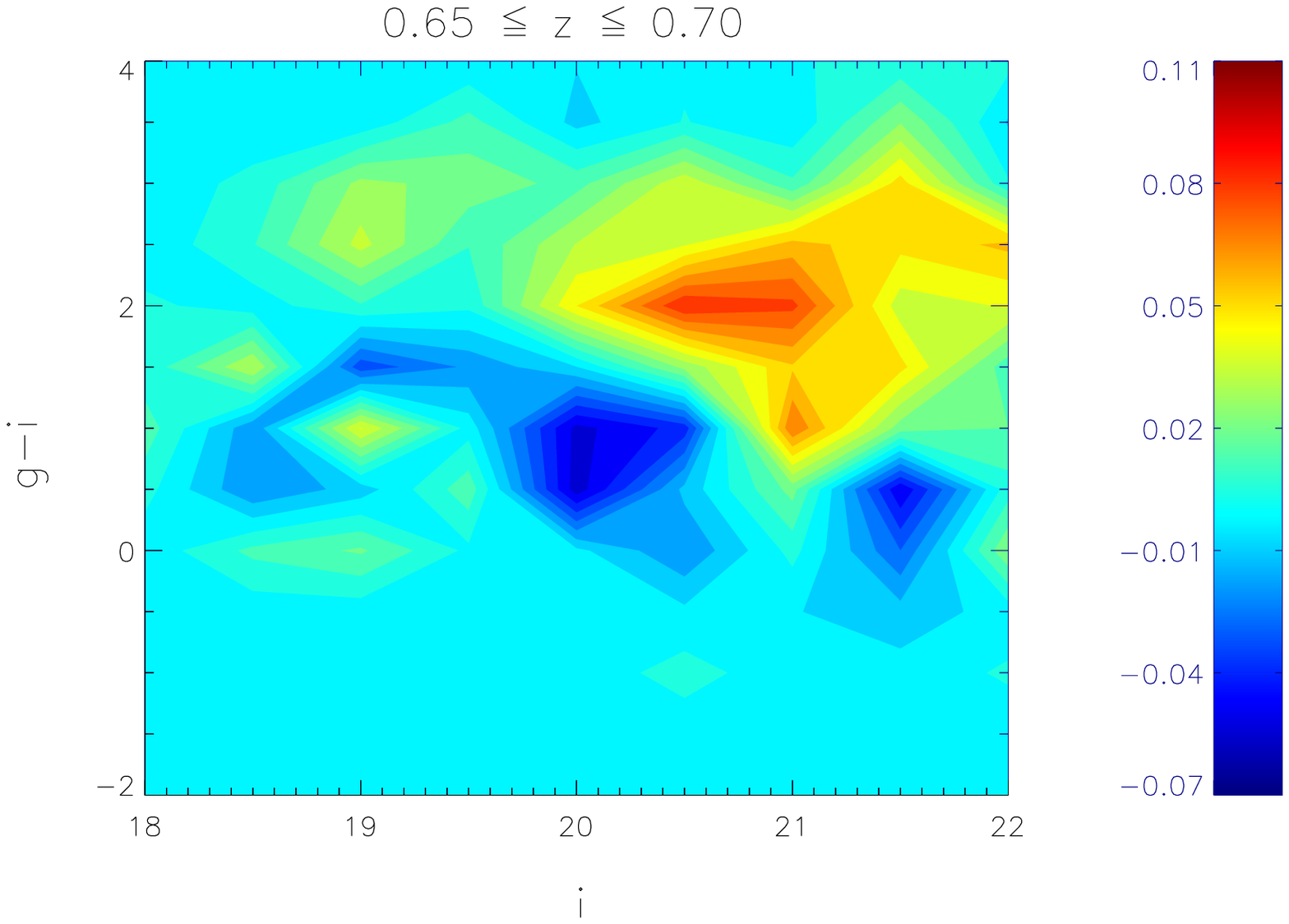} 
	\caption{{Colour-Magnitude diagrams for background subtracted SDSS galaxies. The maps show a clear trend in colour-magnitude space that resembles a cluster red sequence.}}\label{fig:cm_plots}
\end{figure*}

To look for the colour-magnitude relation of the groups and clusters, $i$-magnitude vs $(g-i)$-colour maps were made for both field galaxies and cluster$+$field galaxies in each SDSS square. The C-M maps were then stacked by cluster redshift in redshift slices as seen in Table-\ref{tab:zrange}. Finally, the stacked field galaxy maps were subtracted from the stacked cluster galaxy maps taking into account the relative areas. Fig.\ref{fig:cm_plots} shows the background subtracted C-M diagrams for SDSS galaxies. The maps show a clear trend in colour-magnitude space that resembles a cluster red sequence. The cluster red sequence is an observational property whereby cluster galaxies are more red than field galaxies at the same redshift. As with the plots in fig.\ref{fig:nz_plots} this trend is visible out to $z\sim 0.6$ and similarly indicates that the structures are genuine.

\subsection{Mass estimates}
\begin{figure*}
	\centering
	\includegraphics[width=8cm]{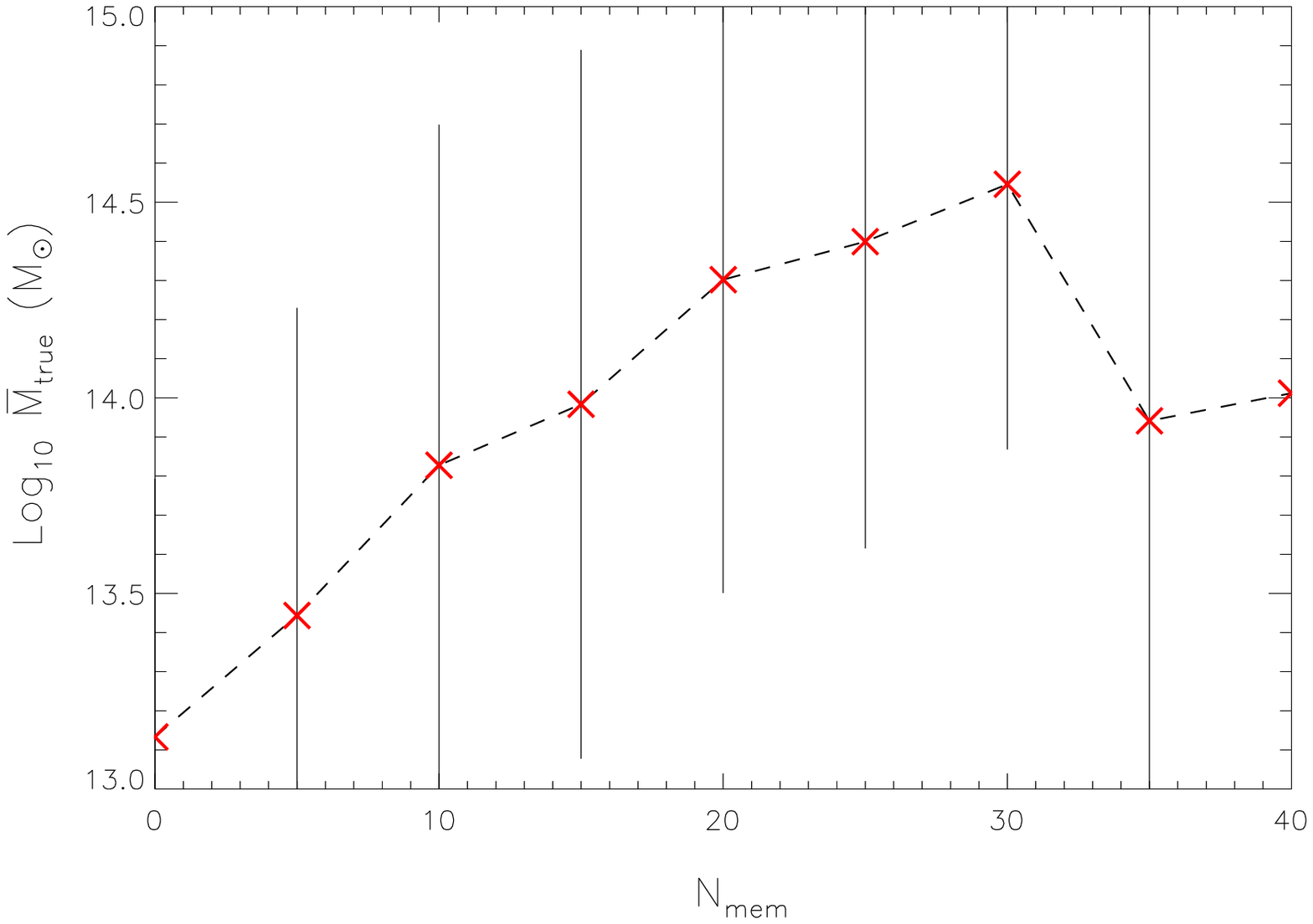} 
	\includegraphics[width=8cm]{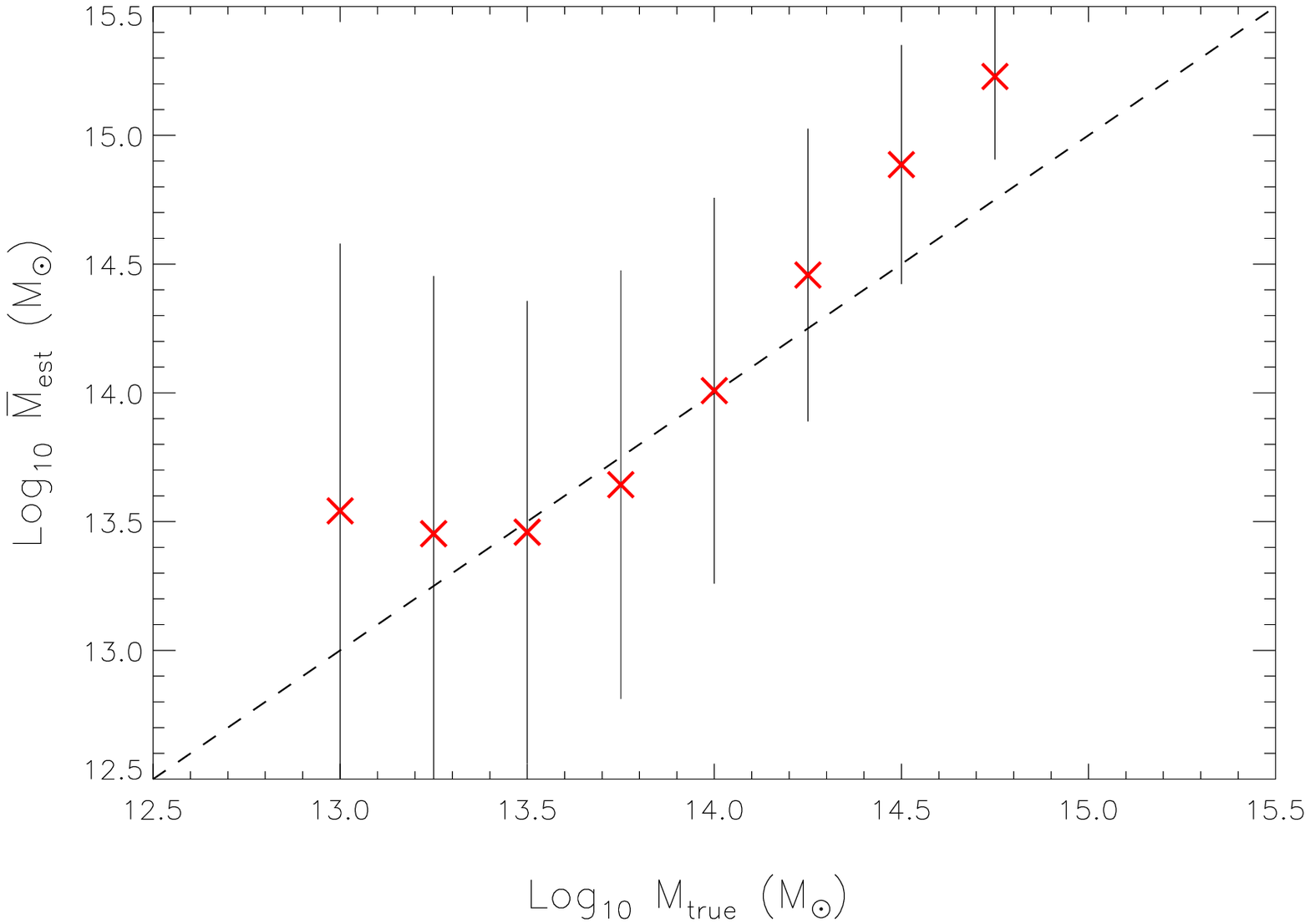} 
	\caption{{Average true cluster mass, $\overline{M}_{true}$, as a function of cluster richness, $N_{mem}$ (left panel). Average estimated mass, $\overline{M}_{est}$, as a function of true mass (right panel). Error bars are the standard error of the mean.}\label{fig:mass_v_n}}
\end{figure*}
In order to calculate masses for our sample of groups and clusters, we compare mass estimates calculated using equation \ref{eq:sigmavir} with the true masses of the 2SLAQ simulation haloes. Masses were calculated for the DFoF catalogue produced from the 2SLAQ mock with linking parameters $R_{friend}(z=0.5)=0.87$ Mpc $h^{-1}$ and $v_{friend}(z=0.5)=900$ kms$^{-1}$ using the cluster velocity dispersions (hereafter $M_{est}$). The membership matching code described in section 5.1 was then implemented to assign to each cluster a mock halo mass (hereafter $M_{true}$). 

Fig.\ref{fig:mass_v_n} shows the average true cluster mass, $\overline{M}_{true}$, as a function of cluster richness, $N_{mem}$, and the average estimated mass, $\overline{M}_{est}$, as a function of true mass. The error bars show the standard error of the mean. The left panel of this plot shows the expected trend between cluster richness and mass. The right panel compares the matched halo masses to those calculated from the cluster velocity dispersions according to equation \ref{eq:sigmavir}. The deviations from the $x = y$ trend are principally the result of contaminating galaxies in the richness estimates and the small number of high mass clusters detected. At the low mass range clusters are composed of only 3 or 4 LRGs, hence their velocity dispersions are unreliable and one object that is either missing of interloping in the cluster will have a large impact on the cluster velocity dispersion and hence the mass estimate. At the high mass range cluster velocity dispersions should be more reliable, but there are very few objects that contribute to the average mass estimate. In addition to this, in the high mass range, when all the galaxies in an individual mock halo have been detected perfectly by the DFoF code, the cluster mass from the velocity dispersion tends to be overestimated by around $10^{0.4}\textrm{M}_\odot$. This is, however, comparable with the size of the error bars in the high mass range of fig.\ref{fig:mass_v_n}.

The range of cluster masses is a good match to known masses of massive clusters. Therefore, from this analysis we can assume that mass estimates made with the real 2SLAQ cluster velocity dispersions will be approximately representative of their true physical masses with a large amount of uncertainty, particularly in the low mass range. This uncertainty could be reduced with a larger sample of clusters and by including all cluster galaxies, rather than just the LRGs, to calculate the velocity dispersions.

\subsection{Clipping}
In section 5.1 we chose to produce a catalogue that optimised the completeness, while maintaining the highest possible purity. This choice resulted in a catalogue that was 98$\%$ complete and 52 $\%$ pure. Fig. \ref{fig:pure_test} shows that clipping groups and clusters with the lowest richnesses will improve the purity, but will reduce the completeness. This plot also shows that the purity is much higher for lower values of $R_{friend}(z=0.5)$, which would correspond to structures that have smaller radial sizes. Therefore, it appears that the purity is mainly affected by the size and richness of the groups and clusters.

In an attempt to improve the purity while preserving the completeness, several new catalogues were made by clipping out groups and clusters with sizes larger than some threshold for a given richness from the original 2SLAQ mock cluster catalogue. Table \ref{tab:clips} shows the total number of groups and clusters with 3 and 4 members for various radial cuts. R$_{clip}$ is the size threshold in Mpc $h^{-1}$ above which all groups and clusters will be removed, N$_{tot}$ is the total number of groups and clusters for a given R$_{clip}$, N$_{true}$ is the number of groups and clusters that are matched to 2SLAQ mock haloes for a given R$_{clip}$ and the ratio N$_{true}$/N$_{tot}$ gives a measure for the purity for a given R$_{clip}$. As can been seen in the table, when no cuts are made groups of 3 and 4 members are only 38$\%$ and 57$\%$ pure respectively. Above richness of 4 all groups and clusters are above 80$\%$ pure. Therefore, the majority of the contamination in the cluster catalogue arises from these small groups.

Taking only groups and clusters with sizes less than 0.11 Mpc $h^{-1}$ for N$_{mem}$ = 3 and less than 0.49 Mpc $h^{-1}$ for N$_{mem}$ = 4 produces a new catalogue which is 94$\%$ complete and 88$\%$ pure. These values were chosen because they provide the best improvement to the purity with minimal change to the completeness. Thus, by clipping out clusters with a low richness ({\it i.e.} few members), but relatively large sizes, we can improve the purity by around 36$\%$, while reducing the completeness by only 4$\%$.

Applying this same clipping procedure to the `real' 2SLAQ groups and clusters, we can separate the catalogue into `gold' and `silver' samples. Where the gold clusters are those that pass the clipping procedure and therefore are the most likely to be genuine. The silver clusters fail the clipping procedure and may still be genuine, but the probability is lower. Out of the 313 total 2SLAQ groups and clusters, 70 are gold and the remaining 243 are silver. The gold sample groups and clusters have an average velocity dispersion of $\overline{\sigma}_{v}=585.28$ kms$^{-1}$ and an average size of $\overline{R}_{clt}=0.90$ Mpc $h^{-1}$, while the silver sample groups and clusters have an average velocity dispersion of $\overline{\sigma}_{v}=430.47$ kms$^{-1}$ and an average size of $\overline{R}_{clt}=0.53$ Mpc $h^{-1}$. The average properties of the gold sample are larger than those of the full (gold$+$silver) sample as a large fraction of the small structures have been removed.

It should be noted that, although we are confident that our mock is a good representation of the 2SLAQ galaxies, it is difficult to interpret how well this clipping procedure will translate to the real 2SLAQ catalogue. Therefore we do not remove the silver sample clusters, rather we assign to them a lower likelihood of being genuine than the gold sample clusters. The final 2SLAQ cluster catalogue contains an additional column that identifies each cluster as belonging to the gold or silver samples with `G' or `S' respectively. In fig.\ref{fig:vel_d} and \ref{fig:size_d} the groups and clusters that form part of the gold sample are highlighted with gold asterisks.

\begin{table}
\caption{Radial Clipping: R$_{clip}$ is the size threshold in Mpc $h^{-1}$ above which all groups and clusters will be removed, N$_{tot}$ is the total number of groups and clusters for a given R$_{clip}$, N$_{true}$ is the number of groups and clusters that are matched to 2SLAQ mock haloes for a given R$_{clip}$ and the ratio N$_{true}$/N$_{tot}$ gives a measure for the purity for a given R$_{clip}$.}
\begin{tabular}{|c|c|clc|}
	\hline
R$_{clip}$ (Mpc $h^{-1}$) & N$_{tot}$ &  N$_{true}$ &  N$_{true}$/N$_{tot}$ \\
	\hline
	\hline
\multicolumn{4}{|c|}{N$_{mem}$ = 3} \\	
	\hline
- & 17382 & 6741 & 0.388 \\
0.60 & 13998 & 6739 & 0.481 \\
0.49 & 11359 & 6739 & 0.593 \\
0.39 & 9367 & 6739 & 0.719 \\
0.28 & 7998 & 6739 & 0.842 \\
0.21 & 7399 & 6738 & 0.911 \\
0.14 & 6993 & 6708 & 0.959 \\
{\bf 0.11} & {\bf 6723} & {\bf 6563} & {\bf 0.976} \\
0.07 & 5789 & 5727 & 0.989 \\
	\hline
\multicolumn{4}{|c|}{N$_{mem}$ = 4} \\	
	\hline
- & 5927 & 3381 & 0.571 \\
0.70 & 4561 & 3306 & 0.725 \\
0.60 & 3987 & 3131 & 0.785 \\
{\bf 0.49} & {\bf 3362} & {\bf 2862} & {\bf 0.851} \\
0.39 & 2966 & 2697 & 0.909 \\
0.32 & 2728 & 2582 & 0.946 \\
0.25 & 2577 & 2505 & 0.972 \\
0.18 & 2433 & 2415 & 0.992 \\
0.11 & 2116 & 2108 & 0.996 \\
	\hline
\end{tabular}
\label{tab:clips}
\end{table}

\subsection{Cluster Images}
\begin{figure*}
	\centering
	\includegraphics[width=8cm, height=5cm]{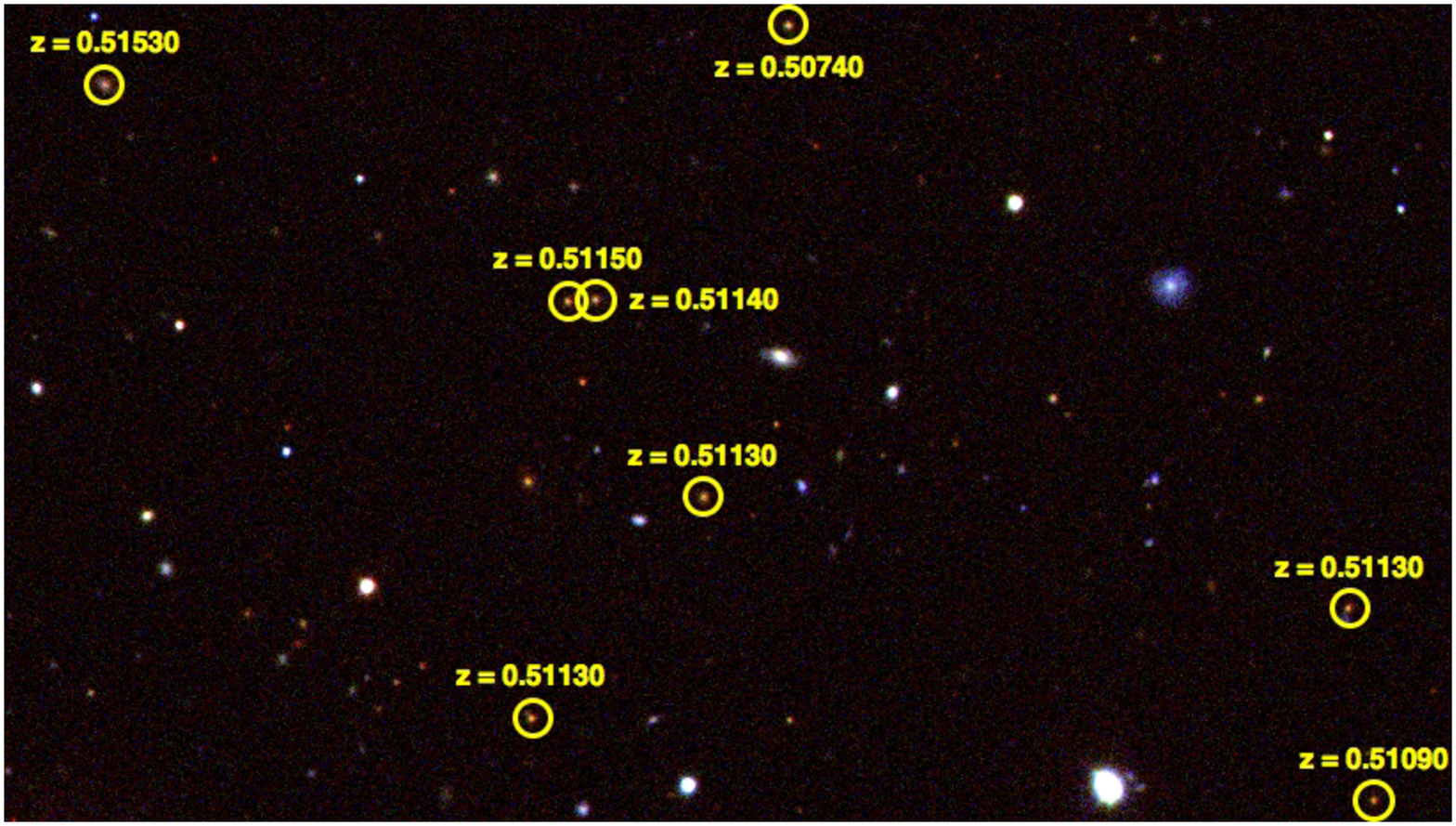} 
	\includegraphics[width=8cm, height=5cm]{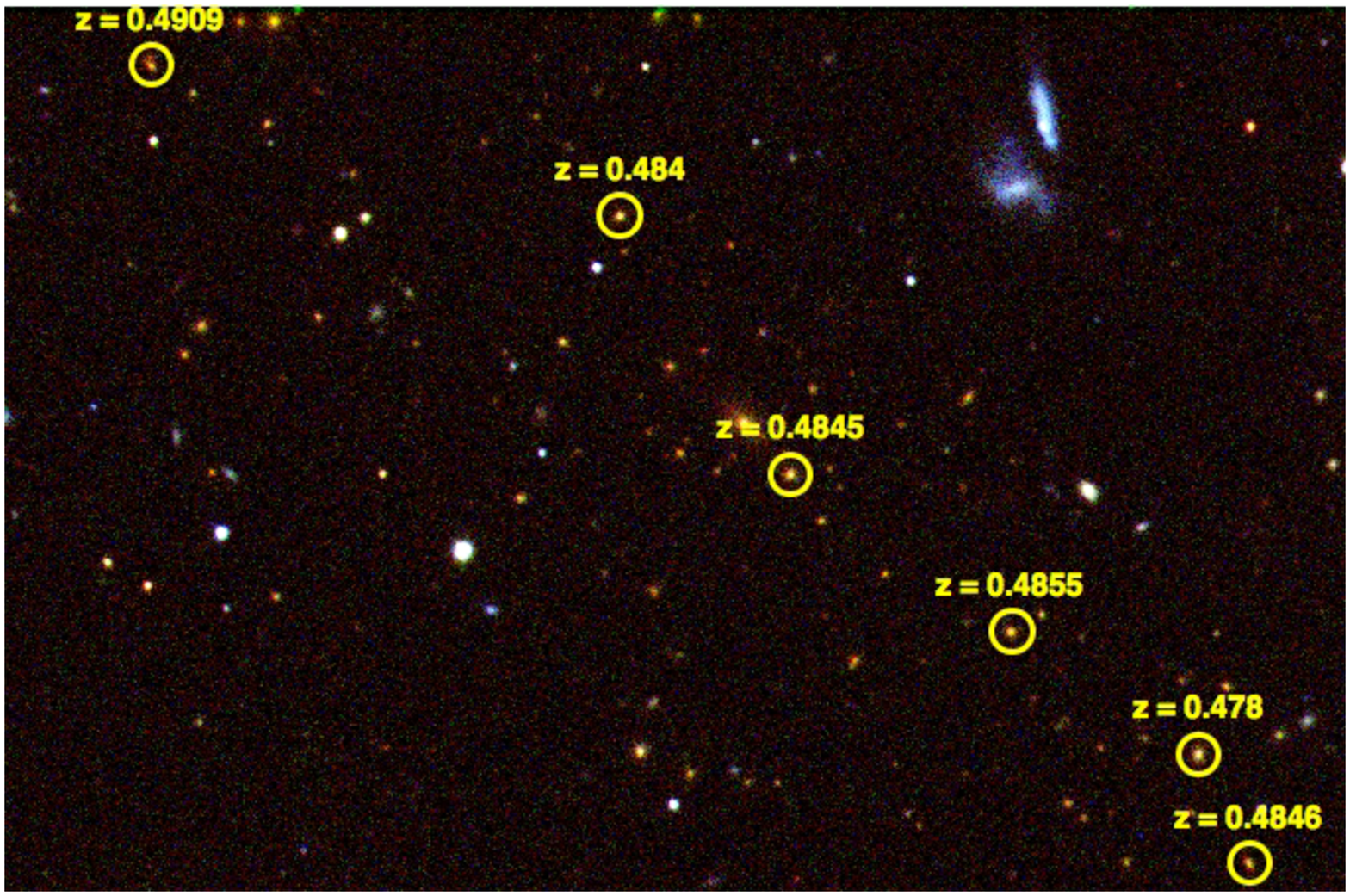} 
	\includegraphics[width=8cm, height=5cm]{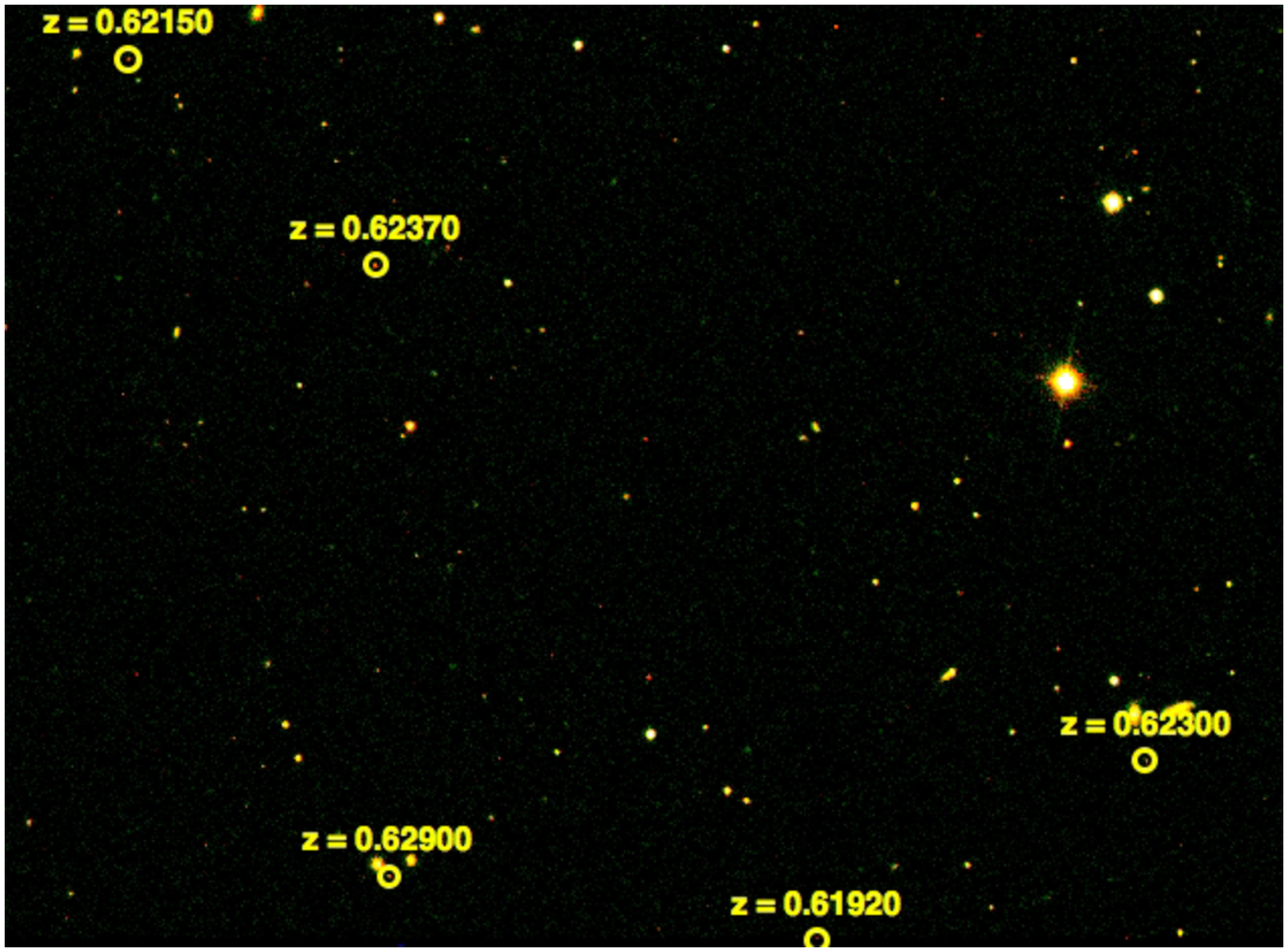} 
	\includegraphics[width=8cm, height=5cm]{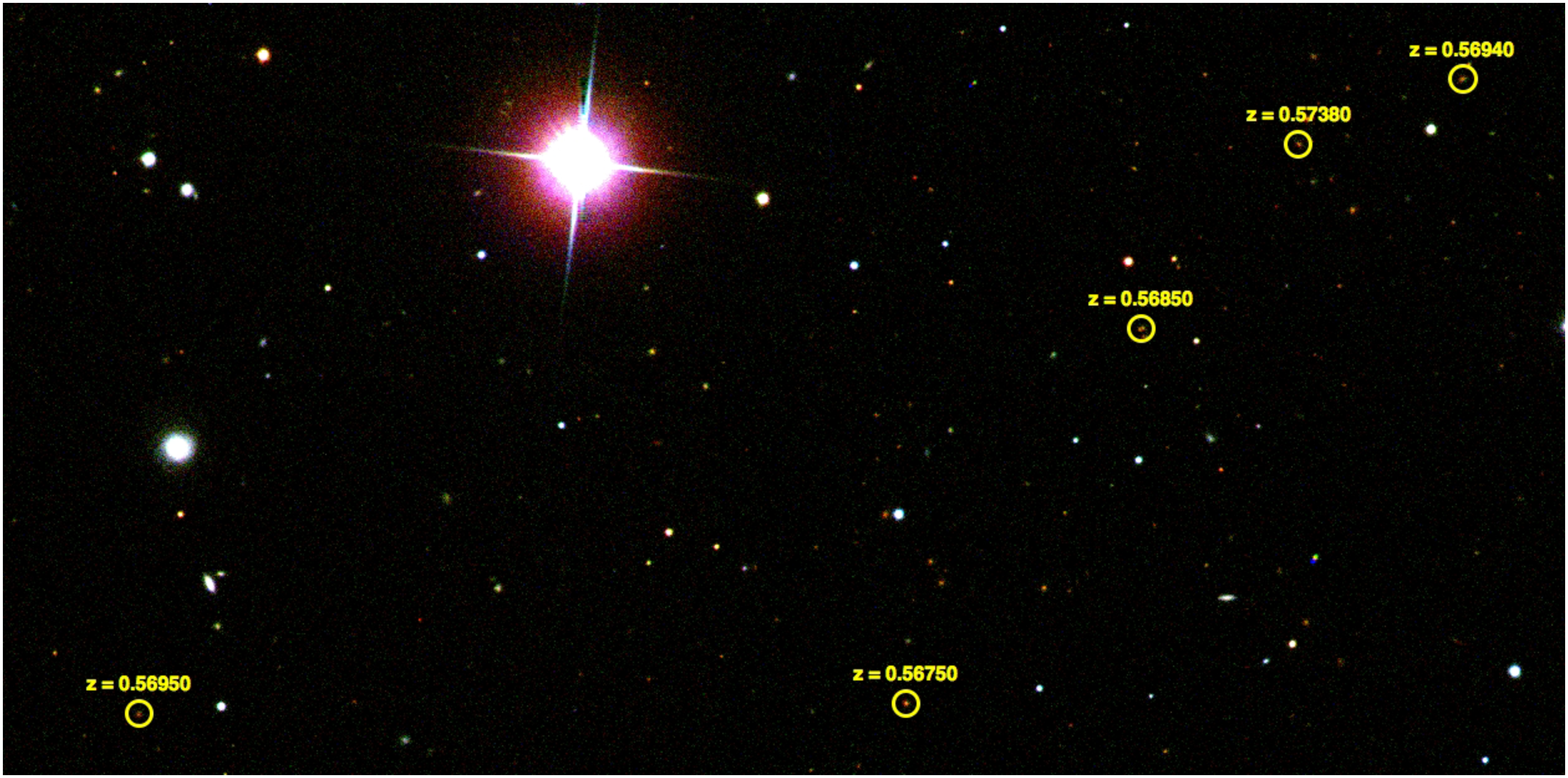} 
	\includegraphics[width=8cm, height=3cm]{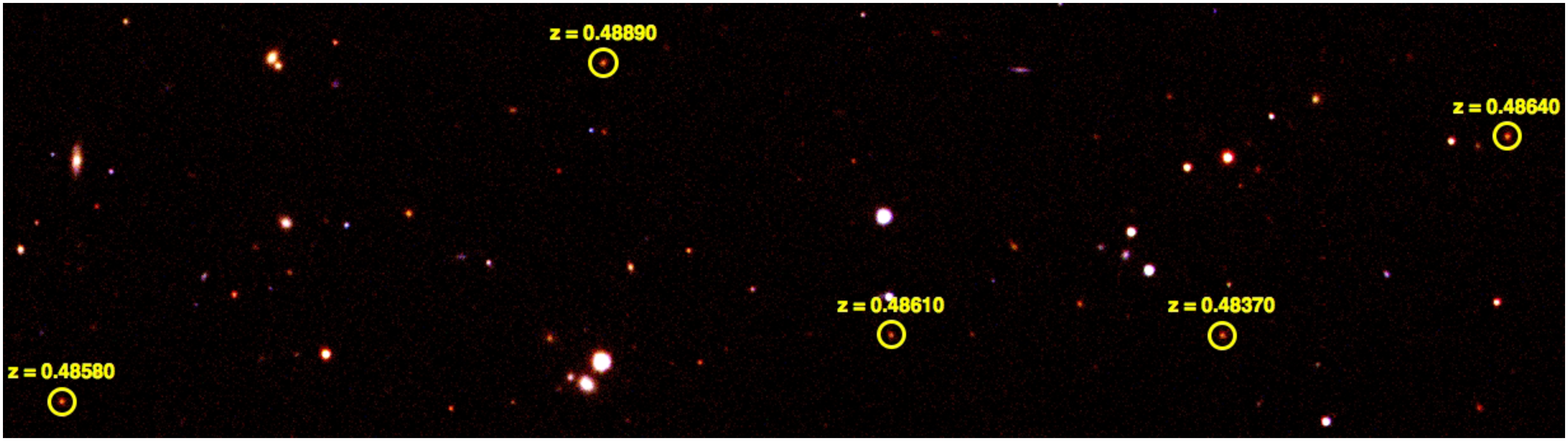} 
	\caption{{SDSS \emph{g, r} and \emph{i}-bands colour images of field around CL$\_$008 (top left panel), CL$\_$204 (top right panel), CL$\_$122 (middle left panel), CL$\_$024 (middle right panel) and CL$\_$038 (bottom panel). Yellow circles indicate the location of the cluster member galaxies, which are labeled with their individual redshifts.}}\label{fig:images}
\end{figure*}

Fig. \ref{fig:images} shows SDSS DR7 optical \emph{g, r} and \emph{i}-band colour images of the regions around the DFoF detected clusters CL\_008, CL\_204, CL\_122, CL\_024 and CL\_038. The yellow circles highlight the positions of the cluster member galaxies, which are labeled with their individual redshifts. These clusters were chosen to sample serval different redshift bins as listed in Table \ref{tab:zrange}.

These images show that the galaxies are distributed in a small area on the sky and are close in redshift space. An important point to notice is that the distribution of cluster LRGs in each image varies from spherical to elongated filamentary structures. This is an advantage of the FoF method, which makes no prior assumptions about the shape of the groups and clusters. CL\_204 (top right panel) in particular shows a rather filamentary distribution of cluster LRGs. The full cluster will be comprised of many galaxy most of which will not be LRGs and therefore the true shape could well be more spherical. Other techniques that have prerequisites on the shape of clusters would not be able to detect this particular object using only the LRGs.

\section{Correlation Function}
The spatial distribution of galaxies and clusters of galaxies contains a wealth of information regarding the underlying cosmological model.  The most widely used method in the literature for condensing this information is to measure the autocorrelation function of the positional data. For our purposes we measure the correlation function simply to compare the statistical spatial distribution of our cluster sample to other derived cluster samples. 

We calculate the two-point correlation function for the 313 2SLAQ  groups and clusters obtained previously. Using the 2SLAQ spectroscopic galaxy catalogue we create a random catalogue, which replicates the angular completeness on the sky, this process is handled straightforwardly in the Healpix\footnote{This software is available from http://healpix.jpl.nasa.gov/} package of software. The radial distribution of the random catalogue is obtained from a smooth spline fit to the galaxy redshift distribution, $n(z)$. The smoothing ensures the exclusion of large scale structure \emph{e.g} voids and filaments.

The correlation function is calculated with the Landy and Szalay estimator \citep{LandySzalay:93}: 
\begin{equation}
\xi = \frac{DD - 2DR + RR}{RR}
\label{eq:landy}
\end{equation}
\noindent where $DD$, $DR$ and $RR$ are the number of pairs of points in the data, $D$, and random, $R$, catalogues. The number of pairs are calculated in 14 bins, equally separated in log space, from $r=5-90$ Mpc $h^{-1}$. 

\begin{figure*}
	\centering
	\includegraphics[width=8cm]{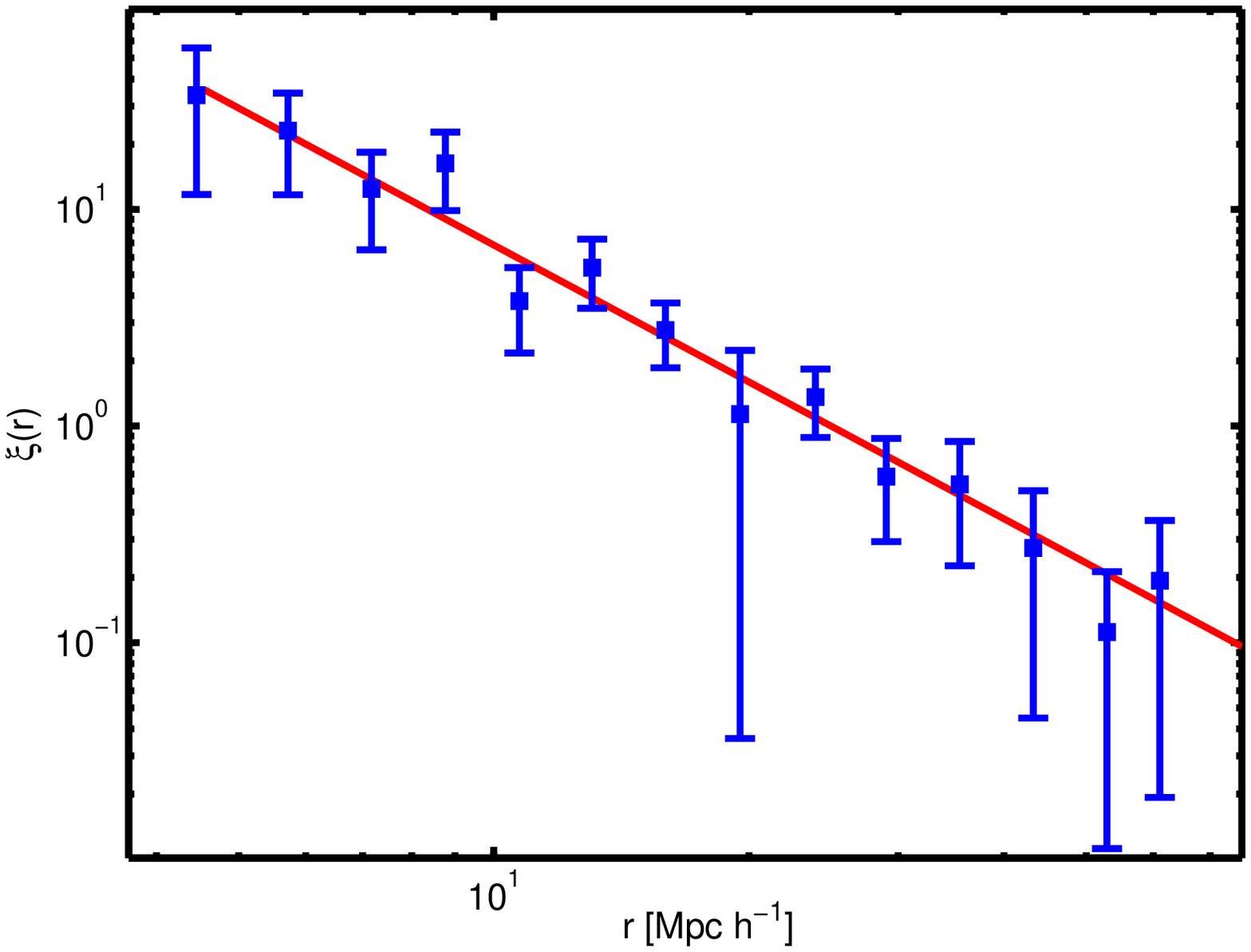} 
	\includegraphics[width=8cm,height=5.8cm]{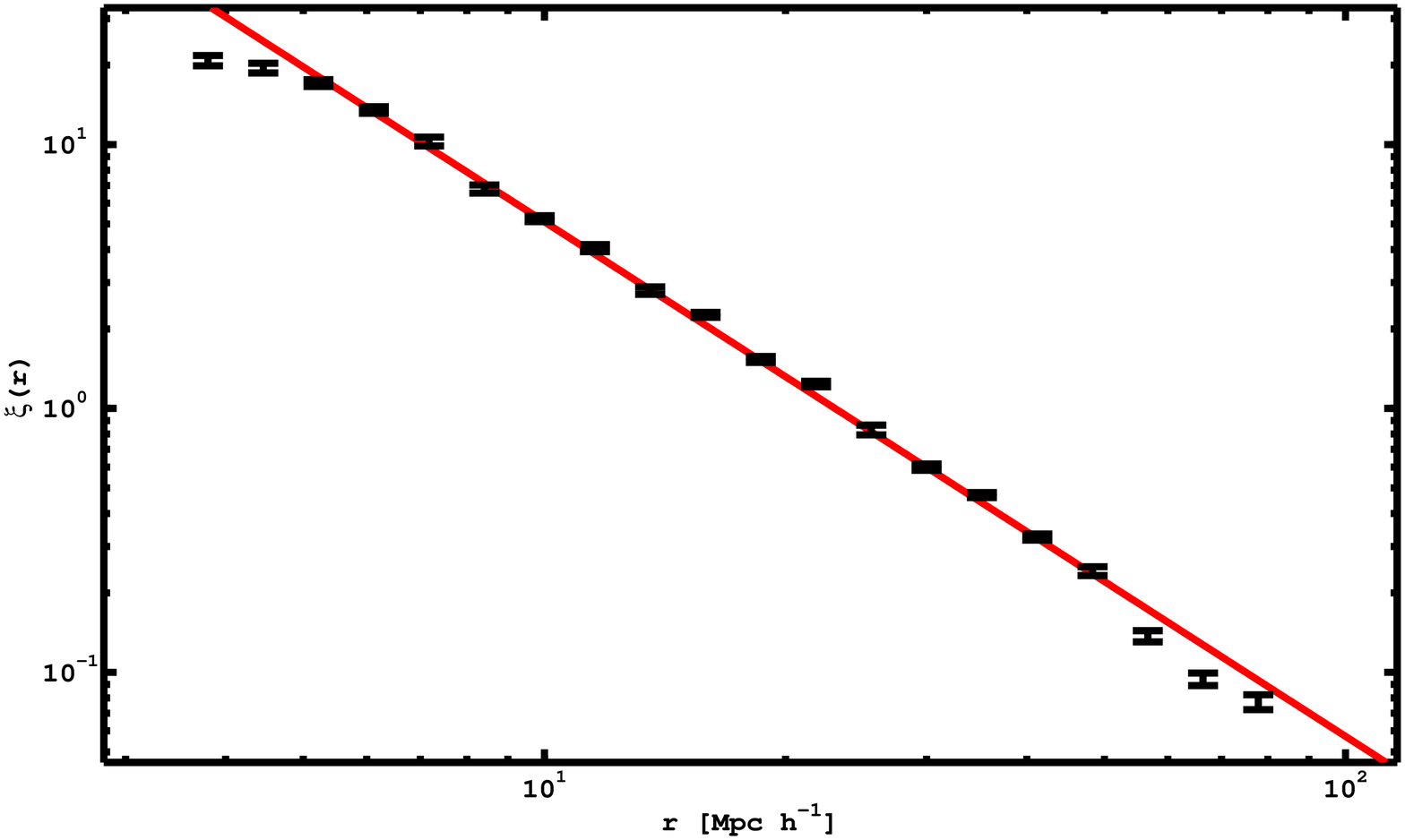}
	\caption{{Left panel: Two-point correlation function for the groups and clusters found in the 2SLAQ catalogue using the DFoF code (blue squares with error bars) and best-fit power law slope (red solid line). Right panel: Two-point correlation function for the mock halo catalogue (black squares with error bars) and best-fit power law slope (red solid line). $\xi_{(R)}$ was measured using the estimator of \citet{LandySzalay:93}. }\label{fig:cluster_2pcf}}
\end{figure*}

The correlation function is usually represented as a power law, $\xi(r)=(r/r_0)^{\gamma}$, thus making a comparison with other works quite straight forward. In the left panel of fig.\ref{fig:cluster_2pcf} we plot the two-point correlation function of the 2SLAQ clusters (blue squares with error bars) and a best-fit power-law slope (red solid line). The error bars are estimated using the jackknife method, which involves dividing the survey into $N$ sections with equal area or volume. The variance and mean are estimated from $N$ measures of our statistic. Each measurement is performed on the survey with region $i$ removed, where $i=1,...,N$. 

The jackknife estimate of the variance is \citep{Lupton:93}: 
\begin{equation}
\sigma^2_\xi(r_i)=\frac{N_{jk}-1}{N_{jk}}\sum^{N_{jk}}_{j=1}[\xi_j(r_i)-\overline{\xi}(r_i)]^2 
\end{equation}
\noindent where $N_{jk}$ is the number of Jackknife samples used and $r_i$ represents a single bin in our statistic, $\xi$. 
In this analysis we set $N_{jk}=20$ sample.

The power law model is best fit with the parameters $r_0=24\pm4$ Mpc $h^{-1}$ and $\gamma=-2.1 \pm 0.2$, and the value of the reduced chi-squared is $\chi^2_{red} = 0.94$.

The right panel in fig.\ref{fig:cluster_2pcf} shows the two-point correlation function of the mock haloes (black squares with error bars) and a best-fit power-law slope (red solid line). The correlation of the mock haloes is best fit with parameters $r_0=23.05\pm0.72$ Mpc $h^{-1}$ and $\gamma=-1.95\pm0.05$, which are perfectly consistent with the parameters found for the 2SLAQ groups and clusters. The error bars are measured using the jackknife method as before.

Although the errors are much larger for the real 2SLAQ data, this analysis indicates that the catalogue of groups and clusters found within 2SLAQ using the DFoF code shows the correct level of clustering with respect to the mock halo catalogue. In addition, significant amounts of impurity and/or incompleteness in the cluster catalogue would suppress the correlation length as impurity adds randomly placed structure and incompleteness removes correlated structure. Thus, as both the real and mock 2SLAQ data have consistent correlation lengths, this implies that the amount of incompleteness and impurity in the 2SLAQ cluster catalogue is not significant.

The correlation length of the 2SLAQ LRG sample was calculated by \citet{Ross:07}, $7.45\pm0.35$  Mpc $h^{-1}$ and by  \citet{Sawangwit:09}, $7.5\pm0.04$ Mpc $h^{-1}$, however this value is expected to be larger for clusters of galaxies. Comparing the $r_0$ value obtained with those found using other low redshift cluster samples shows good agreement, {\emph e.g} $r_0 = 26.0 \pm 4.5$ Mpc $h^{-1}$ \citep{Borgani:99}, $19.4\leq r_0 \leq23.3$ Mpc $h^{-1}$ \citep{Miller:99}, $18.8 \pm 0.9$ Mpc $h^{-1}$ \citep{Collins:00}. See \citet{Nichol:01} for a good review of $r_0$ values for various cluster samples. The clustering length found here is also consistent with that expected for $\Lambda$CDM, $22\leq r_0 \leq27$ Mpc $h^{-1}$ \citep{Colberg:00}.

\label{sec:2pcfs}

\section{Conclusions}
We have written an optical cluster finding algorithm based on that of \citet{HuchraGeller:82}. Our dynamic friends-of-friends (DFoF) code uses a linking length that compensates for selection effects by changing size depending on the surface number density of galaxies at a given redshift. 

We produced a mock catalogue, which is representative of the 2SLAQ catalogue, using the Horizon $4\pi$ simulation and the HOD prescription of \citet{blake:08} in order to determine the linking parameters,  $R_{friend}(z)$ and $v_{friend}(z)$, in the DFoF code. The code was run with various combinations of the two parameters producing a set of distinct catalogues. We wrote a membership matching code in order to determine the completeness and purity of the resulting catalogues relative to the original mock haloes. Based on this analysis we chose values of $R_{friend}(z=0.5) = $ 0.87 Mpc $h^{-1}$ and $v_{friend}(z=0.5) = $ 900 kms$^{-1}$, which correspond to a catalogue that is 98$\%$ complete and 52$\%$ pure with the largest number of groups and clusters possible. Running the DFoF code with these values, we produced a catalogue of 313 groups and clusters containing 1152 member galaxies. The galaxy groups and clusters have an average velocity dispersion of $\overline{\sigma}_{v}=467.97$ kms$^{-1}$ and an average size of $\overline{R}_{clt}=0.78$ Mpc $h^{-1}$.

We tested the validity of our catalogue by obtaining SDSS galaxies in a 1deg$^2$ region around each cluster centre. We then subtracted the background signal from all the SDSS galaxies within 1 Mpc $h^{-1}$ of the cluster centre and stacked the results in bins according to cluster redshift. Each of these bins was examined in SDSS photometric redshift space and colour-magnitude space. This analysis shows that our groups and clusters are reliable out to $z \sim 0.6$. The discrepancies beyond this range may be owing to the small number of groups and clusters detected at higher redshifts. Also, there may be some contamination effects in the histograms from the photometric redshift errors in the SDSS data.

We produce mass estimates for our catalogue using the cluster velocity dispersions according to equation \ref{eq:sigmavir}. We tested the reliability of this approach by comparing mass estimates for clusters found in the 2SLAQ mock with the true masses of the mock haloes to which they are matched. We find that the mass estimates of our mock catalogue groups and clusters are for the most part a good fit to the true halo masses, however with large error bars. The deviations seen are the result of contaminating galaxies in the richness estimates and the small number of high mass clusters detected. In general we see that the range of cluster masses is a good match to known masses of massive clusters. 

We analysed optical SDSS \emph{g, r} and \emph{i}-band colour images of a selection of the our clusters, which span different redshifts. We observe that the galaxy members are distributed in a small region of space, on the sky and in redshift. We also see an overdensity of red galaxies around the cluster centres. This is strong evidence that the clusters are genuine. The distribution of LRGs in the clusters varies from spherical to elongated filamentary structures. This highlights an advantage of the percolation method in that it makes no prior assumptions about the cluster shapes, which allows us to detect some structures that other methods may not.

We test different clipping procedures on the 2SLAQ mock group and cluster sizes. This analysis indicated that the majority of the contamination in the cluster catalogue was attributed to groups with 3 or 4 members. We found that by clipping out clusters with R$_{clt} >$ 0.11 Mpc $h^{-1}$ for N$_{mem}$ = 3 and R$_{clt} >$ 0.49 Mpc $h^{-1}$ for N$_{mem}$ = 4 improves the purity from 52$\%$ to 88$\%$, while reducing the completeness by only 4$\%$, in the 2SLAQ mock catalogue. By applying this procedure to the real 2SLAQ group and cluster catalogue, we separated the clusters into `gold' and `silver' samples. Where the gold samples consists of clusters that passed the clipping procedure and are therefore the most reliable and the silver sample consists of clusters that failed the clipping procedure and may still be genuine, however less reliable.  Out of the 313 total 2SLAQ groups and clusters, 70 are gold and the remaining 243 are silver.

Finally, we test the two-point correlation function of our cluster catalogue. We find a best-fitting power law model , $\xi(r)=(r/r_0)^{\gamma}$, with parameters $r_0=24 \pm 4$ Mpc $h^{-1}$ and $\gamma=-2.1 \pm 0.2$. The value of the reduced chi-squared is $\chi_{red}^2 = 0.94$. These values are consistent with those of the mock halo catalogue and are in good agreement with those in literature \citep{Nichol:01}.

Future surveys such as the Dark Energy Survey (DES), Euclid and Planck will images millions of galaxies across the whole sky. An abundance of photometric data will be obtained for each of these objects, however it will not be possible to obtain spectroscopic data for all of them. Therefore, it is important to develop reliable cluster finding techniques that utilise the photometric data that will be available. Photometric redshifts, for example, provide a useful way to probe the properties of galaxies along the line of sight when spectroscopic data is not present. All of the results presented in the paper will be compared to those found in Farrens et al. (in prep), which will examine galaxy clusters in the 2SLAQ catalogue using photometric redshifts. This comparison will test the reliability of the DFoF code to detect structures using photometric data.

\section*{Acknowledgements}
The authors would like to thank Nic Ross for data and assistance with 2SLAQ correlation functions, Prof. Ofer Lahav for his advice and support, Ignacio Ferreras and  Marcus Vin\'{i}cius Costa Duarte for help regarding k-corrections,  the DES clusters working group for discussions, John Deacon, Fabrizio Sidoli and Dugan Witherick for technical support, and Magda Vasta for helpful suggestions and tips. The authors also acknowledge the use of the UCL Legion High Performance Computing Facility, and associated support services, in the completion of this work. SF and CGS acknowledge support received from the Science and Technology Facilities Council and FBA acknowledges support received from the Royal Society via URF.
\bibliography{aamnem99,biblist}

\appendix

\section{Cluster Catalogue}
The catalogue contains 313 clusters with the following properties: identifier, number of galaxy members (or richness), right ascension, declination, redshift, velocity dispersion (in kms$^{-1}$), radial size (in Mpc $h^{-1}$), estimated mass (in log$_{10}$ M$_{\odot}$) and sample to which it belongs (G: gold ; S: silver). A sample of the first 10 clusters is shown in table-\ref{tab:catalogue}.
\onecolumn
\begin{table}
\caption{Cluster Catalogue Sample}
\begin{tabular}{|c|c|clc|c|c|clclclcl}
	\hline
N$\bar{o}$ & ID & N$_{mem}$ & RA & Dec &  z & $\sigma_{v}$ (kms$^{-1}$) & R$_{clt}$ (Mpc $h^{-1}$) & log$_{10}$ M$_{est}$ (M$_{\odot}$) & Sample\\
	\hline
1 & CL\_0008 & 011 & 152.85693 & -0.00323 & 0.51165 & 0562.51 & 02.08 & 14.3051 & G\\
2 & CL\_0031 & 010 & 210.59379 & -0.55691 & 0.44007 & 1178.60 & 02.00 & 15.2916 & G\\
3 & CL\_0060 & 009 & 035.42705 & -0.22677 & 0.59094 & 0784.58 & 01.74 & 14.7136 & G\\
4 & CL\_0140 & 008 & 342.39005 & -0.44521 & 0.49785 & 0846.21 & 01.27 & 14.8415 & G\\
5 & CL\_0204 & 007 & 158.86053 & -0.02893 & 0.48510 & 1108.13 & 00.65 & 15.1969 & G\\
6 & CL\_0038 & 007 & 191.56150 & -0.51722 & 0.48679 & 0511.13 & 01.66 & 14.1882 & G\\
7 & CL\_0051 & 007 & 190.76950 & +0.03930 & 0.46779 & 0786.90 & 01.26 & 14.7564 & G\\
8 & CL\_0122 & 007 & 161.24470 & -0.30396 & 0.62371 & 0851.02 & 01.85 & 14.8094 & G\\
9 & CL\_0064 & 006 & 202.19576 & -0.35692 & 0.52837 & 0381.83 & 00.73 & 13.7950 & G\\
10 & CL\_0024 & 006 & 343.37305 & -0.49057 & 0.56982 & 0591.89 & 01.52 & 14.3531 & G\\
	\hline
\end{tabular}
\label{tab:catalogue}
\end{table}

\end{document}